\documentclass[preprint,superscriptaddress,showpacs,amsmath,amssymb,aps,pra]{revtex4-1}
\usepackage{times}
\usepackage{amssymb}
\usepackage{graphicx}
\usepackage{dcolumn}
\usepackage{dsfont}
\usepackage{bm}
\usepackage{subfigure}
\usepackage[ruled]{algorithm2e}
\usepackage{hyperref}
\usepackage{appendix}
\usepackage{enumerate}
\usepackage{amsmath}
\usepackage{booktabs}
\makeatletter
\renewcommand*\env@matrix[1][\arraystretch]{%
  \edef\arraystretch{#1}%
  \hskip -\arraycolsep
  \let\@ifnextchar\new@ifnextchar
  \array{*\c@MaxMatrixCols c}}
\makeatother
\newtheorem{theorem}{Theorem}[section]
\newtheorem{lemma}[theorem]{Lemma}

\newtheorem{corollary}[theorem]{Corollary}

\newcommand{\qihao}{\fontsize{3.05pt}{\baselineskip}\selectfont}

\newtheorem{example}{Example}[section]

\newcommand{\p}{\ensuremath{{\bf Proof.\ }}}

\hypersetup{colorlinks=true, citecolor=blue, urlcolor=blue, linkcolor=blue}

\begin{document}
\title{Mutually unbiased maximally entangled bases from difference matrices}
\author{Yajuan Zang}
\affiliation{School of Mathematical Sciences, Capital Normal University, Beijing, 100048, China}
\author{Zihong Tian}
\affiliation{School of Mathematical Sciences, Hebei Normal University, Shijiazhuang, 050024, China}
\author{Hui-Juan Zuo}
\affiliation{School of Mathematical Sciences, Hebei Normal University, Shijiazhuang, 050024, China}

\author{Shao-Ming Fei}
\email{feishm@cnu.edu.cn}
\affiliation{School of Mathematical Sciences, Capital Normal University, Beijing, 100048, China}

\begin{abstract}
Based on maximally entangled states, we explore the constructions of mutually unbiased  bases in bipartite quantum systems. We present a new way to construct  mutually unbiased bases by  difference matrices in the theory of combinatorial designs.
In particular, we establish $q$ mutually unbiased bases with $q-1$ maximally entangled bases and one product basis in $\mathbb{C}^q\otimes \mathbb{C}^q$ for arbitrary prime power $q$. In addition, we construct maximally entangled bases for dimension of composite numbers of non-prime power, such as five maximally entangled bases in $\mathbb{C}^{12}\otimes \mathbb{C}^{12}$ and $\mathbb{C}^{21}\otimes\mathbb{C}^{21}$, which improve the known lower bounds for $d=3m$, with $(3,m)=1$ in $\mathbb{C}^{d}\otimes \mathbb{C}^{d}$.
Furthermore, we construct $p+1$ mutually unbiased bases with $p$ maximally entangled bases and one product basis in $\mathbb{C}^p\otimes \mathbb{C}^{p^2}$ for arbitrary prime number $p$.

\end{abstract}
\parskip=3pt

\maketitle

\section{Introduction}
\label{intro}
The mutually unbiased bases (MUBs) are tightly related to the complementarity of measurements on complementary quantum mechanical observables \cite{Scully}. They are widely utilized in  quantum state tomography \cite{Ivanovi,Wootters}, quantum kinematics \cite{Durt} and quantum error correction codes \cite{Pawlowski}. Moreover, MUBs also play an important role in various tasks in quantum information processing such as quantum key distribution \cite{Cerf}, cryptographic protocols \cite{Brierley, Cerf}, mean king problem \cite{Aharonov}, quantum teleportation and superdense coding \cite{Durt1,Revzen,Sangare}.
Two orthonormal bases $\mathcal{B}_1=\{|\phi_i\rangle\}_{i=0}^{d-1}$ and $\mathcal{B}_2=\{|\psi_j\rangle\}_{j=0}^{d-1}$ of a $d$-dimensional vector space $\mathbb{C}^d$ are called \emph{ mutually unbiased} if
\begin{equation}
|\langle\phi_i|\psi_j\rangle|=\frac{1}{\sqrt{d}},~~~(i,j\in[d]),
\end{equation}
where $[d]=\{0,1,\ldots,d-1\}$. It implies that if a system is in an eigenstate of a certain basis, the measurement outcome with respect to the corresponding MUBs is completely uncertain.

It has been shown that there are no more than $d+1$ MUBs in $\mathbb{C}^d$, which can be attained when $d$ is a single prime power \cite{Ivanovi,Wootters}. For instance, for any prime dimension $d=p$, the eigenstates of the $p+1$ operators,
$Z,~X,~XZ,~XZ^2,\ldots,XZ^{p-1}$, form a full set of $(p+1)$ Heisenberg-Weyl MUBs, where $Z$ and $X$ are the generalized Pauli operators such that $X|j\rangle=|j+1 ~$mod$ ~d\rangle$ and $Z|j\rangle=\omega_d^j|j\rangle$ for the given $d$-dimensional basis $|j\rangle$, $j\in[d]$.
For $d = p^n$, the unitary operators $X^{k_1}Z^{l_1}\otimes \cdots\otimes X^{k_n}Z^{l_n}$, $k_j,l_j\in [p]$, acting on the Hilbert space $\mathbb{C}^p\otimes \cdots \otimes \mathbb{C}^p$, generate the complete set of MUBs. By partitioning the operators into $d+1$ commuting classes, the common eigenvectors of the operators in each class just constitute one of the bases. Moreover, the $d+1$ classes yield $d+1$ unbiased bases \cite{Hiesmayr}. Nevertheless, when $d$ is a composite number of non-prime power, the maximum number of MUBs of $\mathbb{C}^d$ is still unknown.

Quantum entanglement is considered to be one of the most striking features of quantum mechanics. Especially the maximally entangled states such that whose reduced states are maximally mixed play an important role in quantum information processing. Recent years, the study on mutually unbiased maximally entangled bases (MUMEBs) has attracted much attention \cite{cheng,Liu,Shi,Song,Tao,Xu2}. Let $M(d, d')$ be the maximal cardinality of any set of MUMEBs in $\mathbb{C}^d\otimes \mathbb{C}^{d'}$. Liu et al. constructed $p_1^{\alpha_1}-1$ MUMEBs in $\mathbb{C}^d\otimes \mathbb{C}^{d}$ for $d=p_1^{\alpha_1}\cdots p_s^{\alpha_s}$ with $p_1^{\alpha_1}\leq\cdots\leq p_s^{\alpha_s}$ \cite{Liu}. Xu constructed $2(d-1)$ MUMEBs in $\mathbb{C}^d\otimes \mathbb{C}^{d}$ for $d=p^\alpha$, and min$\{q,M(d,d)\}$ MUMEBs in $\mathbb{C}^d\otimes \mathbb{C}^{qd}$ for $d=p^\alpha$ and $q=p'^{\alpha'}$ \cite{Xu2}. Cheng et al. constructed $2(p_1^{\alpha_1}-1)$ MUMEBs in $\mathbb{C}^d\otimes \mathbb{C}^{d}$  for odd $d=p_1^{\alpha_1}\cdots p_s^{\alpha_s}$ with $p_1^{\alpha_1}\leq\cdots\leq p_s^{\alpha_s}$, and min$\{(p'_1)^{\alpha'_1}+1,M(d,d)\}$ MUMEBs in $\mathbb{C}^d\otimes \mathbb{C}^{kd}$ for odd $d$ and $k=(p'_1)^{\alpha'_1}\cdots (p'_l)^{\alpha'_l}$ with $(p'_1)^{\alpha'_1}\leq\cdots\leq (p'_l)^{\alpha'_l}$,  which improved Liu's and Xu's results \cite{cheng}.

Difference matrix was first introduced by Bose and Bush in combinatorial designs \cite{Bose}. It has been studied primarily as a consequence of their uses in the constructions of orthogonal arrays and Latin squares (see, for example, \cite{Beth,Colbourn}). Difference matrix has also been found useful in the constructions of authentication codes without secrecy \cite{Stinson},  data compression \cite{Korner}, software
testing \cite{Cohen,Cohen1} and general Steiner triple systems related to constant weight codes \cite{Wu}.

In this paper, by using difference matrices we investigate the mutually unbiased bases for maximally entangled bases in bipartite systems. We introduce some basic definitions and facts needed in Sec. \ref{sec:Preliminaries}. In Sec. \ref{MUMEB1}, we present $q$ MUBs with $q-1$ MEBs and one product basis (PB) in $\mathbb{C}^q\otimes \mathbb{C}^q$ for arbitrary prime power $q$. In addition, we present MUMEBs for the dimension of composite numbers of non-prime power such as 5 MUMEBs in $\mathbb{C}^{12}\otimes \mathbb{C}^{12}$, $5$ MUMEBs in $\mathbb{C}^{21}\otimes\mathbb{C}^{21}$. In Sec. \ref{MUMEB2}, we directly establish a connection between a difference matrix and maximally entangled bases. Finally, we provide $p+1$ MUBs with $p$ MEBs and one PB in $\mathbb{C}^p\otimes \mathbb{C}^{p^2}$ for arbitrary prime number $p$. By the way, we also give the corresponding examples in $\mathbb{C}^{12}\otimes \mathbb{C}^{12}$, $\mathbb{C}^{3}\otimes \mathbb{C}^{6}$ and $\mathbb{C}^{3}\otimes \mathbb{C}^{9}$. Conclusions and discussions are given in Sec. \ref{conclusion}.

\section{Difference matrices and Latin squares}
\label{sec:Preliminaries}
In this section, we introduce the concepts of difference matrix and Latin square. As well as we give the connection between a difference matrix and a Latin square. Meanwhile we also prove some preliminary results needed for the rest sections.

Let $(G,\odot)$ be a group of order $d$. A $(d,N,\lambda)$-\emph{difference matrix} denoted by $(d,N,\lambda)$-DM is an $N\times \lambda d$ matrix
$M=(m_{il})$ with entries from $G$, so that for different $ i,j\in [N]$, the multiset
$\{m_{il}\odot m_{jl}^{-1}: l\in [\lambda d]\}$  contains every element of $G$ $\lambda$ times. When $G$ is abelian, typically
additive notation is used, so that difference $m_{il}-m_{jl}$ is employed. In this paper, we default that the index of the row of a DM begins at 0.

A difference  matrix is \emph{normalized} if all entries in its first row are all the identity element of the group $G$. Obviously, any difference matrix can be written as normalized by $m_{0,l}^{-1}$ action on any element $m_{i,l}$  in $i$th-row in $M$ for each $ i\in [N]$ and $l\in[\lambda d]$.
Note that if a difference matrix is normalized, then each element occurs exactly $\lambda$ times in arbitrary row except for the $0$th-row. Otherwise, if a difference matrix only consists of rows with each element occurring exactly $\lambda$ times, then it can always be written as normalized by adding a row with all identity.
\begin{example}\label{dm12}
Below is a normalized $(12,6,1)$-DM in $\mathbb{Z}_2\times\mathbb{Z}_6$ which will be used to construct MUMEBs and PB in $\mathbb{C}^{12}\otimes \mathbb{C}^{12}$.
\end{example}
\vspace{-0.2cm}\begin{equation}
M=\left(
\begin{array}{cccccccccccc}
00&00&00&00&00&00&00&00&00&00&00&00\\
00&01&02&03&04&05&10&11&12&13&14&15\\
00&03&10&01&13&15&02&12&05&04&11&14\\
00&12&01&15&05&13&03&14&02&11&10&04\\
00&04&15&14&02&11&12&10&13&01&03&05\\
00&10&12&02&11&01&13&15&04&14&05&03
\end{array}
\right).\label{lidm12}
\end{equation}

A \emph{Latin square} of order $d$, denoted by LS$(d)$, is a $d\times d$ array in which each cell contains a single symbol from a $d$-set $S$, such that each symbol occurs exactly once in each row and exactly once in each column.
Let $M=(m_{il})$ be a $(d,N,\lambda)$-DM in a group $(G,\odot)$, and $m_i$ be the multiset formed by the elements of the $i$th-row of $M$, $ i\in [N]$. For any $g\in G$, define
$m_i\odot g=\{m_{il}\odot g: l\in [\lambda d]\}$. Any multiset $m_i\odot g$ is called a $translate$ of the $i$th-row of $M$. Then, define Dev$(M)$ to be the collection
of all $d$ translates of all rows of $M$, i.e. $Dev(M)=\{m_{i}\odot g: g\in G,  i \in [N]\}$.
Dev($M$) is called the $development$ of $M$. Actually, when $M$ is normalized and $\lambda=1$, the development of the $i$th-row of $M$ forms a Latin square of order $d$ for $1\leq i\leq N-1$.

Concerning the development, let us consider a non-normalized $(q,q+1,q)$-DM for detail which will be needed for rest sections.
Let $q$ be a prime power and  $\mathbb{F}_q$ be a finite filed of order $q$. Denote $\mathbb{F}_q=\{\alpha_0=0,\alpha_1=1,\alpha_2,\ldots,\alpha_{q-1}\}$ and $\mathbb{F}^*_q=\{\alpha_1=1,\alpha_2,\ldots,\alpha_{q-1}\}$ the multiplicative group of $\mathbb{F}_q$. Then $M$ is a $(q,q+1,q)$-DM  as follows \cite{Colbourn}:
\begin{eqnarray*}
\qihao
\begin{small}
\setlength{\arraycolsep}{1.2 pt}
 M=\left(
\begin{array}{cccccccccccccc}
0&0&\cdots&0&\cdots&\alpha_i&\cdots&\alpha_i&\cdots&\alpha_i&\cdots&\alpha_{q-1}&\cdots&\alpha_{q-1}  \\
0&1&\cdots&\alpha_{q-1}&\cdots&0&\cdots&\alpha_j&\cdots&\alpha_{q-1}&\cdots&0&\cdots&\alpha_{q-1}   \\
0&1&\cdots&\alpha_{q-1}&\cdots&\alpha_{i}&\cdots&\alpha_{i}+\alpha_j&\cdots&\alpha_{i}+\alpha_{q-1}&\cdots&\alpha_{q-1}&\cdots&\alpha_{q-1}+\alpha_{q-1}   \\
&&&&&&&\cdots&&&\\
0&\alpha_g&\cdots&\alpha_g\alpha_{q-1}&\cdots&\alpha_{i}&\cdots&\alpha_i+\alpha_g\alpha_j&\cdots&\alpha_i
+\alpha_g\alpha_{q-1}&\cdots&\alpha_{q-1}&\cdots&\alpha_{q-1}+\alpha_g\alpha_{q-1}  \\
&&&&&&&\cdots&&&\\
0&\alpha_{q-1}&\cdots&\alpha^2_{q-1}&\cdots&\alpha_{i}&\cdots&\alpha_i+\alpha_{q-1}\alpha_j&\cdots&\alpha_i+\alpha_{q-1}^2&\cdots&\alpha_{q-1}&\cdots&\alpha_{q-1}+\alpha^2_{q-1}
\end{array}
\right),
\end{small}
\end{eqnarray*}
\hspace{2cm}$\underbrace{\hspace{2.5cm}}\ \ \cdots \hspace{0.2cm}\underbrace{\hspace{5cm}}\hspace{0.4cm} \cdots \hspace{0.2cm}\underbrace{\hspace{3cm}}$
\begin{equation}\label{dm2}
\vspace{-0.2cm}\hspace{0.3cm}P_0 \hspace{4.4cm}P_i \hspace{4.8cm}P_{q-1}
\end{equation}
with the partitions $P_0,P_1,\ldots,P_{q-1}$, where $P_i$ is a $(q+1)\times q$ submatrix for $i\in [q]$. Actually, if $r>1$, each translate of the $r$th-row of $M$ forms a Latin square. Namely, for any fixed $\alpha_t\in \mathbb{F}_q$, we can define a Latin square of order $q$ with $L(i,j)=\alpha_i+\alpha_{r-1}\alpha_j+\alpha_t$, $i,j\in [q]$, where $i$ ($j$) is the index of row (column) in Latin square. On the other hand, there contain exactly $q$ intersections in any two translates respectively from the $r$th-row and the $r'$th-row of $M$ for $r\neq r'$. Moreover each symbol in $\mathbb{F}_q$ appears exactly once in these intersections. Especially, when $q$ is a prime number, we have the following results, see  Appendix \ref{sec:AppA} for the proof.

\begin{lemma}\label{qq}
When $p$ is a prime number, for positive $r\neq r'$ the $p$ intersection points of any two translates respectively from the $r$th-row and the $r'$th-row in the difference matrix $M$
(\ref{dm2}) are different and have the properties below:

\begin{enumerate}
	\item if $r=1$ and $r'=2$, the $p$ intersection points just appear at the development of  submatrix $P_i$ for some $i\in [p]$;
	\item if $r=1$ and $r'>2$,  the $p$ intersection points are respectively from the development of submatrix $P_i$ for each $i\in [p]$. Moreover, the difference of two adjoining intersections is a fixed number;
	\item if $r, r'>1$,  the $p$ intersection points are respectively from the development of submatrix $P_i$  with the same column index $j$ for each $i\in [p]$. Moreover,
the intersections just constitute a shift of $(0,1,\ldots,p-1)$.
\end{enumerate}
\end{lemma}

Musto introduced the concept of weak orthogonal  Latin squares (WOLS) in order to construct MUMEBs \cite{Musto}.  Given a pair of Latin squares $L$ and $K$ with entries $l_{ij}$ and $k_{ij}$ respectively, they are \emph{weak orthogonal}  when for all different  $ i,j\in [d]$, if there exists a unique $ s\in [d]$ such that $l_{is}=k_{js}$. A set of $t\geq 2$ Latin squares of order $d$,  say $L^1, L^2,\dots, L^t$,  is said to be \emph{mutually weak
orthogonal}, denoted by $t$-MWOLS($d$), if $L^i$ and $L^j$ are weak orthogonal for all $1 \leq i < j\leq t$.

Actually,  difference matrix has a close relationship with weak orthogonal  Latin squares.

\begin{lemma}\label{dm}
Corresponding to every normalized $(d,N,1)$-DM, there is an $(N-1)$-MWOLS$(d)$.
\end{lemma}

\noindent \p Assume $M=(m_{ij})$ is a normalized $(d,N,1)$-DM based on group $G$. Let $L^i$ be the development of the $i$th-row of $M$, $1\leq i\leq N-1$, i.e.,
\begin{equation}
L^{i}=\left(
\begin{array}{cccc}
m_{i1}\odot g_0 & m_{i2}\odot g_0&\cdots &m_{id}\odot g_0 \\
m_{i1}\odot g_1 & m_{i2}\odot g_1&\cdots &m_{id}\odot g_1 \\
\cdot&\cdot&\cdots&\cdot\\
m_{i1}\odot g_{d-1} & m_{i2}\odot g_{d-1}&\cdots &m_{id}\odot g_{d-1}
\end{array}
\right),
\end{equation}
where $g_s\in G, s\in[d]$. Obviously, $L^1,L^2,\cdots$ and $L^{N-1}$ are LS($d$)s.
For any $i\neq j,s,s'\in [d]$, if $m_{ix}\odot g_s=m_{jx}\odot g_{s'}$, then $m_{jx}^{-1}\odot m_{ix}=g_{s'}\odot g_s^{-1}$. Since $M$ is a difference matrix, so there exactly exists one solution of $ x\in [d]$. Therefore, $L^i$ and $L^j$ are weak orthogonal for any $1\leq i\neq j\leq N-1$.
$\Box$

\begin{corollary}\label{dmc}
For a normalized $(d,N,1)$-DM, any translate of the $r$th-row and any translate of the $r'$th-row of the DM just intersect at one point for arbitrary  $1\leq r\neq r'\leq N-1$.
\end{corollary}

\section{Constructions for MUMEBs in $\mathbb{C}^d\otimes \mathbb{C}^d$}
\label{MUMEB1}
In this section, we construct mutually unbiased maximally entangled bases  and product basis  from a $(d,N,1)$-DM in $\mathbb{C}^d\otimes \mathbb{C}^d$.
A bipartite pure state $|\Psi\rangle$ in systems $A$ and $B$ is said to be maximally entangled in $\mathbb{C}^d\otimes \mathbb{C}^{d'}$ ($d\leq d'$)
if and only if for an arbitrary given orthonormal complete basis $\{|\phi_i\rangle\}_{i=0}^{d-1}$ of the subsystem $A$,
there exists an orthonormal basis $\{|\psi_j\rangle\}_{j=0}^{d'-1}$ of the subsystem $B$ such that
$$
|\Psi\rangle=\frac{1}{\sqrt{d}}\sum\limits_{i\in[d]}|\phi_i\rangle\otimes|\psi_i\rangle.
$$
A basis of $\mathbb{C}^d\otimes \mathbb{C}^{d'}$ ($d\leq d'$) constituted by maximally entangled states is called a
maximally entangled basis (MEB).

A complex Hadamard matrix of order $d$ is a $d\times d$ matrix $H$
with entries $H_{ij}$ such that
\begin{eqnarray}
&\hspace{0.35cm}  |H_{ij}|=1,\  \ \  \ \ \ \ \ \ \ \ \ \ \ \hspace{0.8cm} H_{ij}H_{ij}^*=1,\\
&H H^{\dagger}=dI_d, \ \ \ \ \ \ \ \ \ \ \ \ \sum\limits_{p\in[d]}H_{ip}H^*_{jp}=d\delta_{ij},\\
&H^{\dagger} H=dI_d, \ \ \ \ \ \ \ \ \ \ \ \ \sum\limits_{p\in[d]}H_{pi}H^*_{pj}=d\delta_{ij},
\end{eqnarray}
where $I_d$ is the identity matrix of order $d$. Obviously, $\hat{F}_d\equiv\sqrt{d}F_d$ is a complex Hadamard matrix with $F_d$ corresponding with the $d\times d$ Fourier matrix.

\begin{lemma}\label{lsmub2}
Given $N$ indexed families of complex Hadamard matrices $H_0,H_1,\ldots,H_{N-1}$ of order $d$. With respect to a normalized $(d,N,1)$-DM,
there exist $N$ MUBs with $N-1$ MEBs and a PB in $\mathbb{C}^d\otimes \mathbb{C}^d$.
\end{lemma}

\noindent \p Assume that $M$ is a normalized $(d,N,1)$-DM, and $L^r=(l^r_{ij})$ is the development of the $r$th-row of $M$ for $1\leq r\leq N-1$. Define
\begin{eqnarray}
&A^0_{i,j}=\frac{1}{\sqrt{d}}|i\rangle \otimes \sum\limits_{k\in[d]}|k\rangle\langle k|H_{0}|j\rangle, i,j \in[d],\\
&A^r_{i,j}=\frac{1}{\sqrt{d}}\sum\limits_{k\in[d]}|l^r_{ik}\rangle\otimes |k\rangle\langle k|H_r|j\rangle, i,j\in[d],1\leq r\leq N-1.
\end{eqnarray}
It is easy to check that $\{A^0_{i,j}:i,j \in [d]\}$ is a set of PB and $\{A^r_{i,j}: i,j \in[d]\}$ is a set of MEB for each $1\leq r\leq N-1$. Next we show the mutual unbiasedness.

For any $1\leq r\leq N-1$, we have
\begin{eqnarray*}
&\left|(A^{0}_{i,j},A^r_{i',j'})\right|=\frac{1}{d}\left|\langle k|H_{r}|j\rangle^*\langle k|H_{0}|j'\rangle\right| \\
&\hspace{-1.1cm}=\frac{1}{d}.
\end{eqnarray*}

For any $1\leq r\neq r' \leq N-1$, by Corollary \ref{dmc}  we have
\begin{eqnarray*}
\left|(A^{r}_{i,j},A^{r'}_{i',j'})\right|&=\frac{1}{d}\left|\sum\limits_{k,k'\in[d]}\langle l^r_{ik} | l^{r'}_{i'k'}\rangle\langle k|k'\rangle\langle k|H_{r}|j\rangle^*\langle k'|H_{r'}|j'\rangle\right|\\
&\hspace{-1.5cm}=\frac{1}{d}\left|\sum\limits_{k\in[d]}\langle l^r_{ik} | l^{r'}_{i'k}\rangle \langle k|H_{r}|j\rangle^*\langle k|H_{r'}|j'\rangle\right|\\
&\hspace{-3.6cm}=\frac{1}{d}\left|\langle k|H_{r}|j\rangle^*\langle k|H_{r'}|j'\rangle\right|\\
&\hspace{-7.2cm}=\frac{1}{d},
\end{eqnarray*}
which completes the proof. $\Box$

\begin{example} \label{12}
As an example of the Lemma \ref{lsmub2}, we have that there exist 5 MUMEBs and a PB from a normalized $(12,6,1)$-DM in $\mathbb{C}^{12}\otimes \mathbb{C}^{12}$,
see Appendix \ref{sec:AppB} for the detailed expressions.
\end{example}

There have been fruitful results on the researches of difference matrices, not only for different orders, but also for abelian and non-abelian groups \cite{Buratti,Drake,Evans,Ge,Pan,Pan0,Pan1}. Below are some useful properties of difference matrices.

\begin{lemma}\label{jielun}(\cite{Buratti,Drake,Evans,Ge,Pan})
\begin{enumerate}
	\item For any prime power $q$, there exists a $(q,q,1)$-DM and a $(q,q+1,q)$-DM.
	\item Suppose $G$ is any group of order $d$, then there exists a $(d,p,1)$-DM over $G$ where $p$  is the smallest prime dividing $d$.
	\item A $(d,4,1)$-DM exists if and only if $d\geq 4$ and $d \not\equiv 2$ {\rm(mod 4)}.
	\item A $(d, 5, 1)$-DM exists over $\mathbb{Z}_d$  if and only if $d$ is odd and
$d\not \in \{3,9\}$, except possibly when $d = 9p$ and $p$ is an odd prime other than $3, 5, 7,
11, 13, 17,19, 23, 29, 31,$ or $109$.
    \item If $m \neq 3, 9$ is odd and not of the form $9n$, $n$ not divisible by
$3, 5, 7, 11, 13, 17, 23, 29,\\ 31$, or $109$, then there exists a $(16m,5,1)$-DM over Dihedral group $D(8m)$.
\end{enumerate}
\end{lemma}

\begin{lemma}\label{jielun2}(\cite{Abel,Colbourn,Johnson,Kufeld,Shen})
There exists a $(12,6,1)$-DM, $(21,6,1)$-DM, $(24,8,1)$-DM, $(33,6,1)$-DM, $(39,6,1)$-DM,
$(48,9,1)$-DM, $(51,6,1)$-DM, $(57,7,1)$-DM, $(75,8,1)$-DM and $(273,16,1)$-DM .
\end{lemma}

Combing Lemma \ref{lsmub2}-\ref{jielun2}, we have the main results of this section.

\begin{theorem}\label{jielunmub}
\begin{enumerate}
	\item For any prime power $q$, $M(q,q)\geq q-1$.
	\item For any group $G$ of order $d$, $M(d,d)\geq p-1$, where $p$ is the smallest prime dividing $d$.
	\item If $d\geq 4$ and $d \not\equiv 2$ {\rm(mod 4)}, then $M(d,d)\geq 3$.
	\item If $d$ is odd and
$d\not \in \{3,9\}$, then $M(d,d)\geq 4$ except possibly when $d = 9p$ and $p$ is an odd prime other than $3, 5, 7,
11, 13, 17,19, 23, 29, 31$, or $109$.
 \item If $m \neq 3, 9$ is odd and not of the form $9n$, $n$ not divisible by
$3, 5, 7, 11, 13, 17, 23, 29,\\ 31$, or $109$, then  $M(16m,16m)\geq 4$.
\end{enumerate}
\end{theorem}

\begin{theorem}\label{jielun1}
$M(12,12),M(21,21),M(33,33),M(39,39),M(51,51)\geq 5$; $M(57,57)\geq 6$; $M(24,24),M(75,75) \geq 7$; $M(48,48)\geq 8$; and $M(273,273)\geq 15$.
\end{theorem}

Particularly, we compare the $M(d,d)$ in the above theorem with other known results for some small $d$ with $d=3m$ in Table 1.
\begin{table}[htbp]
\caption{\small  Lower bound on M($d,d$) of MUMEBs for some small $d$ with $d=3m$, $(3,m)=1$. $a$ is the lower bound value $2(p^{\alpha}-1)$ when $d$ is odd \cite{cheng} and $p^\alpha-1$ when $d$ is even  \cite{Liu}, where $p^\alpha$ is the minimal prime power factor of $d$; $b$ is the lower bound value $(p^{2\alpha}-1)/2$ when $d$ is odd and $min\{3(2^t-1), (p^{2\alpha}-1)/2\}$ when $d$ is even \cite{Shi}, where $p^\alpha$ is the minimal odd prime power factor of $d$ and $t$ is the maximal positive integer such that $2^t|d$; $c$ is the lower bound in Theorem \ref{jielun1} in this paper.}\label{eqtable}
\renewcommand\arraystretch{0.8}
\setlength{\tabcolsep}{14pt}
\centering
\begin{tabular}{ccccccccccc}
\toprule[2pt]
$d$ & 12 & 21 &24 &33 & 39 &48 &51 &57 &75 &273  \\
\midrule[1pt]
$M(d,d)^a$ & 2 & 4 &2 &4 & 4 &2 &4 &4 &4 &4  \\
$M(d,d)^b$ & 4 & 4 &4 &4 & 4 &4 &4 &4 &4 &4   \\
$M(d,d)^c$ & 5 & 5 &7 &5 & 5 &8 &5 &6 &7 &15  \\
\bottomrule[2pt]
\end{tabular}
\end{table}

\section{Constructions for MUMEBs in $\mathbb{C}^p\otimes \mathbb{C}^{p^2}$}
\label{MUMEB2}
Similar to the case of $\lambda=1$, for $\lambda\geq 2$ the MEBs in $\mathbb{C}^d\otimes \mathbb{C}^{\lambda d}$ also can be obtained from a normalized $(d,N,\lambda)$-DM. In this section, we introduce a method for constructing MEBs from a normalized $(d,N,\lambda)$-DM and a family of complex Hadamard matrices. Furthermore, we construct mutually unbiased bases with MEBs and one PB in $\mathbb{C}^p\otimes \mathbb{C}^{p^2}$ for arbitrary prime number $p$.

\begin{lemma}\label{qq+1}
Given a normalized $(d,N+1,\lambda)$-DM with $\lambda \geq 2$  and a family of complex Hadamard matrices $H^1_i$ of order $d$ for $1\leq i\leq N-1$ and $H^2_j$ of order $\lambda$  for $ j\in [N]$, then $N-1$ maximally entangled bases and one product basis can be obtained in $\mathbb{C}^d\otimes \mathbb{C}^{\lambda d}$.
\end{lemma}

\noindent \p  Assume $M=(m_{ij})$ to be the $(d,N,\lambda)$-DM after removing the 0th-row of all identity in a normalized $(d,N+1,\lambda)$-DM. Without loss of generality, suppose that the 0th-row in $M$ is $(0,\ldots,0,\ldots,d-1,\ldots,d-1)$ with every element repeated $\lambda$ times.

Let $S^{r}=(s^r_{ij})$ be the development of the $r$th-row of $M$ for $r\in [N]$. Denote
\begin{eqnarray}
  &A^0_{i,j,l}=\frac{1}{\sqrt{\lambda}}|j\rangle\otimes\sum\limits_{
  \{x\in [\lambda d]: s^0_{ix}=j\}} |x\rangle\langle j_x|H^2_0|l\rangle,\\
&A^r_{i,j,l}=\frac{1}{\sqrt{\lambda d}}\sum\limits_{k\in [d]}\langle k|H^1_r|j\rangle|k\rangle\otimes\sum\limits_{
  \{x\in [\lambda d]: s^r_{ix}=k\}} |x\rangle\langle k_x|H^2_r|l\rangle,
\end{eqnarray}
where $i,j\in[d],~ l\in[\lambda],~1\leq r\leq N-1$, $j_x$ or $k_x$ is the rank of element $x$ in the set $\{x\in [\lambda d]: s^r_{ix}=j$ or $k\}$ and here we default the rank starts from zero. Obviously $A^0_{i,j,l}$ is a product state, and  $A^r_{i,j,l}$ is a maximally entangled state under LU operation in $\mathbb{C}^d\otimes \mathbb{C}^{\lambda d}$ for each $1\leq r\leq N-1$, $i,j\in[d]$, $l\in[\lambda]$.

Next we show the orthogonality. Firstly, for any $i,j,i',j'\in [d]$ and $l,l'\in [\lambda]$, we have
\begin{eqnarray}
\hspace{-1cm}(A^0_{i,j,l}, A^0_{i',j',l'} )&=\frac{1}{\lambda}\langle j|j'\rangle \sum\limits_{
  \{x'\in [\lambda d]: s^0_{i'x}=j'\}} \sum\limits_{
  \{x\in [\lambda d]: s^0_{ix}=j\}}\langle x|x'\rangle\langle j_x|H^2_0|l\rangle^* \langle j'_{x'}|H^2_0|l'\rangle \nonumber \\
  &\hspace{-3.3cm}=\frac{1}{\lambda} \sum\limits_{
 \{x\in [\lambda d]: s^0_{ix}=j\}} \langle  j_x|H^2_0|l\rangle^* \langle j_x|H^2_0|l'\rangle\delta_{ii'}\delta_{jj'}\nonumber \\
  &\hspace{-8.5cm}=\delta_{ii'}\delta_{jj'}\delta_{ll'}.
\end{eqnarray}
Secondly, for any $i,j,i',j'\in [d]$, $l,l'\in [\lambda]$ and $1\leq r\leq N-1$, we have
\begin{eqnarray}
\hspace{-1cm}(A^r_{i,j,l}, A^r_{i',j',l'} )&=\frac{1}{\lambda d}\sum\limits_{k\in [d]}\langle k|H^1_r|j\rangle^*\langle k|H^1_r|j'\rangle \sum\limits_{
  \{x\in [\lambda d]: s^r_{ix}=s^r_{i'x}=k\}} \langle k_x|H^2_r|l\rangle^* \langle k_x|H^2_r|l'\rangle \nonumber \\
  &\hspace{-1.9cm}=\frac{1}{\lambda d}\sum\limits_{k\in [d]}\langle k|H^1_r|j\rangle^*\langle k|H^1_r|j'\rangle \sum\limits_{
  h\in [\lambda]} \langle h|H^2_r|l\rangle^* \langle h|H^2_r|l'\rangle\delta_{ii'}\nonumber \\
  &\hspace{-9.2cm}=\delta_{ii'}\delta_{jj'}\delta_{ll'}.
\end{eqnarray}
$\Box$

\begin{example} \label{36}
There exist 4 MEBs and one PB from a non-normalized $(3,5,2)$-DM in $\mathbb{C}^{3}\otimes \mathbb{C}^{6}$. See Appendix \ref{sec:AppC} for the details.
\end{example}

We now use the Heisenberg-Weyl MUBs as the complex Hadamard matrices. In fact, when $p$ is prime, the Heisenberg-Weyl MUBs can be written in matrix form, with $M_0=I$, $M_1=F_p$ and
$M_{i+1}=D^i F_p$ for $i=1,\ldots,p-1$, where $I$ is the identity matrix; $F_p$ is the $p\times p$ Fourier matrix and $D$ is a diagonal matrix \cite{Hiesmayr}. For instance, when $p=3$ or $5$, the complete sets are generated by using $D_3=diag(1,\omega_3,\omega_3)$ and $D_5=diag(1,\omega_5,\omega^4_5,\omega^4_5,\omega_5)$. We show some properties on these Heisenberg-Weyl MUBs below.

\begin{lemma}\label{fd}
\begin{enumerate}
\item $\sum\limits_{i\in [p]}|\langle i|D^m \hat{F}_p|j\rangle^*\langle i|D^{m'}\hat{F}_p|j'\rangle|=\sqrt{p}$. Moreover\\
$\sum\limits_{i\in [p]}|\langle i|I|j\rangle^*\langle i|D^{m}\hat{F}_p|j'\rangle|=\sqrt{p}$ for any $m\neq m'\in [p]$ and $j,j'\in[p]$.
	\item $\langle i|D^m \hat{F}_p|j\rangle^*\langle i|D^{m'}\hat{F}_p|j'\rangle=\langle i|D^{m'-m}\hat{F}_p|j'-j\rangle$ for any $i,j, m, m'\in[p].$
\end{enumerate}
\end{lemma}
\noindent \p \emph{(i)} is obvious since $I$, $D F_p,\ldots,D^{p-1} F_p$ are constituted by $p+1$ MUBs. For the case \emph{(ii)}, suppose $D=diag(\omega_p^{n_0}=1,\omega_p^{n_1},\ldots,\omega_p^{n_{p-1}})$. Then we have
\begin{eqnarray*}
\langle i|D^m \hat{F}_p|j\rangle^*\langle i|D^{m'}\hat{F}_p|j'\rangle&=((\omega_p^{n_{i}})^m\omega_p^{ij})^*((\omega_p^{n_{i}})^{m'}\omega_p^{ij'})\\
&\hspace{-1.5cm}=(\omega_p^{n_{i}})^{m'-m}\omega_p^{i(j'-j)}\\
&\hspace{-1cm}=\langle i|D^{m'-m}\hat{F}_p|j'-j\rangle.
\end{eqnarray*} $\Box$ \newpage

\begin{theorem}
For prime number $p$, there exist $p+1$ MUBs with $p$ MEBs and one PB  in $\mathbb{C}^{p}\otimes \mathbb{C}^{p^2}$.
\end{theorem}

\noindent \p Suppose $H^1_i=D^i \hat{F}_p$  and $H^2_j= \hat{F}_p$ for $1\leq i\leq p$, $ j\in [p]$. Assume that $M$ is a non-normalized $(p,p+1,p)$-DM defined in Eq. (\ref{dm2}) and $S^r=(s^r_{ij})$ is the development of the $r$th-row of $M$ for $1\leq r\leq p$. Let $\{ A^0_{i,j,l}: i,j,l\in[p]\}$ and $\{ A^r_{i,j,l}: i,j,l\in[p]\}$, $1\leq r\leq p$, be the PB and MEBs defined in Lemma \ref{qq+1}, where $d=p$, $N=p+1$ and $\lambda=p$. Namely,
\begin{eqnarray}
  &A^0_{i,j,l}=\frac{1}{\sqrt{p}}|j\rangle\otimes\sum\limits_{
  \{x\in [p^2]: s^0_{ix}=j\}} |x\rangle\langle j_x|\hat{F}_p|l\rangle, \ \ \\
&A^r_{i,j,l}=\frac{1}{p}\sum\limits_{k\in [p]}\langle k|D^r \hat{F}_p|j\rangle|k\rangle\otimes\sum\limits_{
  \{x\in [p^2]: s^r_{ix}=k\}} |x\rangle\langle k_x|\hat{F}_p|l\rangle.
\end{eqnarray}

Next we show the unbiasedness. Firstly  for any $1\leq r \leq p$, we have
\begin{eqnarray*}
|(A^0_{i,j,l},A^r_{i',j',l'})|
&=\frac{1}{p\sqrt{p}}|\langle j|D^{r} \hat{F}_p|j'\rangle| \left |\sum\limits_{
  \{x\in [p^2]: s^0_{ix}=s^r_{i'x}=j\}}\langle j_x|\hat{F}_p|l\rangle^* \langle j_x|\hat{F}_p|l'\rangle\right | \\
 &\hspace{-2.9cm}=\frac{1}{p\sqrt{p}}|\langle j|D^{r} \hat{F}_p|j'\rangle| |\langle j_x|\hat{F}_p|l\rangle^* \langle j_x|\hat{F}_p|l'\rangle |\\
&\hspace{-8.5cm}=\frac{1}{p\sqrt{p}}.
\end{eqnarray*}
Secondly for any $1\leq r\neq r' \leq p$,  by Lemma \ref{qq} and  Lemma \ref{fd}  we have
\begin{eqnarray}
|(A^r_{i,j,l},A^{r'}_{i',j',l'})|&=\frac{1}{p^2} \left| \sum\limits_{k\in [p]}\langle k|D^{r} \hat{F}_p|j\rangle^* \langle k|D^{r'} \hat{F}_p|j'\rangle  \sum\limits_{
  \{x\in [p^2]: s^r_{ix}=s^{r'}_{i'x}=k\}}\langle k_x|\hat{F}_p|l\rangle^* \langle k_x| \hat{F}_p|l'\rangle\right | \nonumber \\
 &\hspace{-2.9cm}=\frac{1}{p^2} \left| \sum\limits_{k\in [p]}\langle k|D^{r} \hat{F}_p|j\rangle^* \langle k|D^{r'}\hat{F}_p|j'\rangle \langle k_x|\hat{F}_p|l\rangle^* \langle k_x|\hat{F}_p|l'\rangle\right |\nonumber \\
 &\hspace{-4.9cm}=\frac{1}{p^2} \left| \sum\limits_{k\in [p]}\langle k|D^{r'-r}\hat{F}_p|j'-j\rangle \langle k_x|\hat{F}_p|l'-l\rangle\right |.  \label{p1}
\end{eqnarray}

If $r=1$ and $r'=2$, since the $i$th-row and the $i'$th-row in $S^0$ and $S^1$ just intersect at $p$ points in the development of $P_h$  for a certain $h$ by Lemma \ref{qq}, one has $k_x=h$ for any $k\in [p]$. Thus $(\langle 0_x|\hat{F}_p|l'-l\rangle,\langle 1_x|\hat{F}_p|l'-l\rangle,\cdots,\langle p-1_{x}|\hat{F}_p|l'-l\rangle)=\omega^{h(l'-l)}(1, 1,\cdots, 1)$. Therefore, Eq. (\ref{p1}) can be further written as \newpage
\begin{eqnarray*}
&~~~\frac{1}{p^2} \left| \sum\limits_{k\in [p]}\langle k|D \hat{F}_p|j'-j\rangle \langle k_x|\hat{F}_p|l'-l\rangle\right | \\
&=\frac{1}{p^2} \left| \sum\limits_{k\in [p]} \omega^{h}\langle k|D \hat{F}_p|j'-j\rangle \langle 0|\hat{F}_p|l'-l\rangle\right |\\
&\hspace{-0.6cm}=\frac{1}{p^2} \left| \sum\limits_{k\in [p]}\langle k|D \hat{F}_p|j'-j\rangle \langle 0|\hat{F}_p|l'-l\rangle\right |\\
&\hspace{-6cm}=\frac{1}{p\sqrt{p}}.
\end{eqnarray*}

If $r=1$ and $r'>2$ with $g>1$, by Lemma \ref{qq} the difference of the two adjoining intersections of the $i$th-row and the $i'$th-row is $(g-1)^{-1}$, i.e., if $k_x=h$ and $k'_x=h+1$, then $k-k'=(g-1)^{-1}$. Hence, we have $0_x-1_x=g-1$ and $k_x-(k+1)_x=g-1$, which implies that $(\langle 0_x|\hat{F}_p|l'-l\rangle,\langle 1_x|\hat{F}_p|l'-l\rangle,\cdots,\langle p-1_{x}|\hat{F}_p|l'-l\rangle)=\omega^{h(l'-l)}(1, w^{-(g-1)(l'-l)},\cdots, w^{-(g-1)(p-1)(l'-l)})$ with $0_x=h$. Therefore, Eq. (\ref{p1}) can be written as
\begin{eqnarray*}
&\hspace{-2.7cm}\frac{1}{p^2} \left| \sum\limits_{k\in [p]}\langle k|D^{r'-r}\hat{F}_p|j'-j\rangle \langle k_x|\hat{F}_p|l'-l\rangle\right | \\
&=\frac{1}{p^2} \left| \sum\limits_{k\in [p]} \omega^{h(l'-l)}\langle k|D^{r'-r}\hat{F}_p|j'-j\rangle \langle k|\hat{F}_p|-(g-1)(l'-l)\rangle\right |\\
&\hspace{-9.5cm}=\frac{1}{p\sqrt{p}}.
\end{eqnarray*}

If $r,r'>1$, with $g\neq g' \in\{1,\ldots,p-1\}$, since the intersections of the $i$th-row and the $i'$th-row just constitute a shift of $(0,1,2,\ldots, p-1)$ by Lemma \ref{qq}, we have $(\langle 0_x|\hat{F}_p|l'-l\rangle,\langle 1_x|\hat{F}_p|l'-l\rangle,\cdots,\langle p-1_{x}|\hat{F}_p|l'-l\rangle)=\omega^{h(l'-l)}(1, w^{(l'-l)},\cdots, w^{(p-1)(l'-l)})$ with $0_x=h$. Therefore, Eq. (\ref{p1}) can also be written as
\begin{eqnarray*}
&\hspace{0.1cm}\frac{1}{p^2} \left| \sum\limits_{k\in [p]}\langle k|D^{r'-r}\hat{F}_p|j'-j\rangle \langle k_x|\hat{F}_p|l'-l\rangle\right | \\
&\hspace{0.7cm}=\frac{1}{p^2} \left| \sum\limits_{k\in [p]} \omega^{h(l'-l)}\langle k|D^{r'-r}\hat{F}_p|j'-j\rangle \langle k|\hat{F}_p|l'-l\rangle\right |\\
&\hspace{-6.7cm}=\frac{1}{p\sqrt{p}}.
\end{eqnarray*}
$\Box$

\begin{example} \label{39}
There exist 3 MUMEBs and one PB from a non-normalized $(3,4,3)$-DM in $\mathbb{C}^{3}\otimes \mathbb{C}^{9}$, see Appendix
\ref{sec:AppD} for the details.
\end{example}

\section{Conclusions}
\label{conclusion}

The Latin square has been used as a quite facilitating tool in the research of MUBs. After Hayashi et al. first presented the proof of the existence of the solutions of the mean king's problem with maximal MUBs in prime power dimensions in terms of the mutually orthogonal Latin squares \cite{Hayashi}, the mutually orthogonal Latin squares have been employed to study MUBs for single systems as well as bipartite systems \cite{Musto,Paterek,Song}. While the difference matrix can not only give rise to mutually orthogonal Latin squares \cite{Abel,Beth,Colbourn,Evans,Johnson,Shen}, but also to the MUMEBs in a more direct way, as shown in this paper. We conjecture that the difference matrix may be also applied to construct MUBs in multipartite systems. Moreover, the difference matrix takes up less storage space than Latin squares, since one row (except for the 0th row) in a normalized difference matrix brings one or $\lambda$ Latin square(s).

We have introduced a new method for constructing MUMEBs via difference matrices in the theory of combinatorial designs. By using difference matrices, we constructed $q$ mutually unbiased bases with $q-1$ MEBs and one product basis in $\mathbb{C}^q\otimes \mathbb{C}^q$ for arbitrary prime power $q$. Furthermore, we constructed MUMEBs for some dimension $d=3m$, where $(3,m)=1$ in $\mathbb{C}^{d}\otimes\mathbb{C}^{d}$ such as 5 MUMEBs in $\mathbb{C}^{12}\otimes \mathbb{C}^{12}$ and $5$ MUMEBs in $\mathbb{C}^{21}\otimes\mathbb{C}^{21}$ etc. (see Table 1), which implied that the bounds of MUMEBs has large probability to be improved especially for bigger dimensions. In addition, we constructed $p+1$  mutually unbiased bases with $p$ MEBs and one product basis in $\mathbb{C}^p\otimes \mathbb{C}^{p^2}$ for arbitrary prime number $p$.
Concerning the existence of MUMEBs there are still many open problems, for example, about the improvement of the lower bound of M$(3m,3m)\geq 4$ for any $d=3m$ with $(3,m)=1$, and that of M$(p,p^2)\geq p$ for prime number.

Recently, many concepts related to the combinatorial designs have been generalized to the field of quantum information, such as quantum Latin squares, quantum orthogonal arrays and quantum Sudoku etc., which have a close relationship with the absolutely maximally entangled (AME) states, $k$-uniform states, orthogonal quantum measurements and MUBs \cite{Goyeneche1,Musto,Paczos,Zang}. Concerning the relationship among difference matrices, Latin squares and orthogonal arrays, it would be also interesting to consider quantum Latin squares, quantum orthogonal arrays or quantum Sudoku from the view of difference matrices. Moreover, a quantum version of difference matrix would also shed new light on the investigation of AME states, $k$-uniform states, orthogonal quantum measurements and the related applications in quantum information processing.

\section*{Acknowledgements}
This work is supported by Beijing Postdoctoral Research Foundation (2022ZZ071), Natural Science Foundation of Hebei Province (F2021205001), NSFC (Grant Nos. 11871019, 12075159, 12171044, 62272208), Beijing Natural Science Foundation (Z190005), Academy for Multidisciplinary Studies, Capital Normal University, the Academician Innovation Platform of Hainan Province, and Shenzhen Institute for Quantum Science and Engineering, Southern University of Science and Technology (No. SIQSE202001).
\appendix
\section{Proof of Lemma \ref{qq}}
\label{sec:AppA}

When $p$ is a prime number, $M$ can be written as
\begin{eqnarray*}\label{dm22}
\qihao
\begin{small}
\setlength{\arraycolsep}{1 pt}
\renewcommand\arraystretch{0.6}
M=\left(
\begin{array}{cccccccccccccc}
0&0&\cdots&0&\cdots&i&\cdots&i&\cdots&i&\cdots&p-1&\cdots&p-1  \\
0&1&\cdots&p-1&\cdots&0&\cdots&j&\cdots&p-1&\cdots&0&\cdots&p-1   \\
0&1&\cdots&p-1&\cdots&i&\cdots&i+j&\cdots&i+p-1&\cdots&p-1&\cdots&p-2   \\
&&&&&&&&\cdots&&&\\
0&g&\cdots&g(p-1)&\cdots&i&\cdots&i+gj&\cdots&i+g(p-1)&\cdots&p-1&\cdots&(g+1)(p-1)  \\
&&&&&&&&\cdots&&&\\
0&p-1&\cdots&(p-1)^2&\cdots&i&\cdots&i+(p-1)j&\cdots&i+(p-1)^2&\cdots&p-1&\cdots&0\\
\end{array}
\right).
\end{small}
\end{eqnarray*}
\hspace{2.2cm}$\underbrace{\vspace{-0.6cm}\hspace{2.5cm}}\ \ \ \cdots \hspace{0.2cm}\underbrace{\hspace{4.3cm}} \hspace{0.6cm} \cdots \hspace{0.2cm}\underbrace{\hspace{2.8cm}}$

\hspace{2.8cm}$P_0$ \hspace{4cm}$P_i$ \hspace{4.3cm}$P_{p-1}$

Case $r=1$ and $r'=2$. For any $s,s'\in [p]$, the equation $j+s=i+j+s'$ has $p$ solutions given by $i=s-s'$ and $j\in [p]$.
Thus the $p$ intersections of the $s$th translate and the $s'$th translate  of the 1st-row and the 2nd-row are different and just located at the development of the submatrix $P_{s-s'}$.
\begin{eqnarray*}
\qihao
\begin{small}
 \setlength{\arraycolsep}{1 pt}
 \renewcommand\arraystretch{0.8}
\begin{array}{lc}
\mbox{}&
\begin{array}{|c|c|c|c|c|c|c|c|c|c|}
\hline
{\color{red}0}&{\color{red}\cdots}&{\color{red}p-1}&{\color{green}0}&{\color{green}\cdots}&{\color{green}p-1}&\cdots&{\color{blue}0}&{\color{blue}\cdots}&{\color{blue}p-1}   \\
\hline
1&\cdots&0&1&\cdots&0&\cdots&1&\cdots&0   \\
\hline
\cdot&\cdots&\cdot&\cdot&\cdots&\cdot&\cdots&\cdot&\cdots&\cdot\\
\hline
p-1&\cdots&p-2&p-1&\cdots&p-2&\cdots&p-1&\cdots&p-2   \\
\hline
\end{array}
\end{array}~~
\begin{array}{lc}
\mbox{}&
\begin{array}{|c|c|c|c|c|c|c|c|c|c|}
\hline
{\color{red}0}&{\color{red}\cdots}&{\color{red}p-1}&1&\cdots&0&\cdots&p-1&\cdots&p-2   \\
\hline
1&\cdots&0&2&\cdots&1&\cdots&{\color{blue}0}&{\color{blue}\cdots}&{\color{blue}p-1}  \\
\hline
\cdot&\cdots&\cdot&\cdot&\cdots&\cdot&\cdots&\cdot&\cdots&\cdot\\
\hline
p-1&\cdots&p-2&{\color{green}0}&{\color{green}\cdots}&{\color{green}p-1}&\cdots&p-2&\cdots&p-3   \\
\hline
\end{array}
\end{array}
\end{small}
\end{eqnarray*}

Case $r=1$ and $r'>2$ with $g>1$. For any $s,s'\in[p]$,  the equation $j+s=i+gj+s'$ has $p$ solutions given by $i=s-s'+(1-g)j$ and $j\in[p]$. Thus the $p$ intersections are different and respectively from the development of the submatrix $P_i$, $i\in[p]$. Suppose $j_0+s=i_0+gj_0+s'$ and $j_1+s=(i_0+1)+gj_1+s'$. Then we have $j_0=(g-1)^{-1}(-i_0+s-s')$ and $j_1=(g-1)^{-1}(-i_0-1+s-s')$. Therefore, the difference of arbitrary two adjoining intersections is $\pm(g-1)^{-1}$.

Case $r,r'>1$, with $g\neq g' \in\{1,\ldots,p-1\}$. For any $s,s'\in[p]$, the equation $i+gj+s=i+g'j+s'$ has $p$ solutions given by $j=(g'-g)^{-1}(s-s')$ and $i\in [p]$. Thus the $p$ intersection points are different and respectively located at the same column index $j$ of the development of submatrix $P_i$ for each $i\in [p]$. Moreover they just constitute a shift of $(0,1,\ldots,p-1)$.

\section{Example \ref{12}: MUMEBs and PB in $\mathbb{C}^{12}\otimes \mathbb{C}^{12}$}
\label{sec:AppB}

Let $L^i$, $i=0,1,...,5$, be the development of the $i$th-row of  difference matrix (\ref{lidm12}) in Example \ref{dm12}  as follows (Here we map $\mathbb{Z}_2\times \mathbb{Z}_6$ onto $\mathbb{Z}_{12}$ with bijection map $f: (i,j)\rightarrow 6i+j$). \newpage
\begin{eqnarray*}
\footnotesize
 \setlength{\arraycolsep}{0.4 pt}
 \renewcommand\arraystretch{0.6}
\begin{array}{|c|c|c|c|c|c|c|c|c|c|c|c|}
\hline
0&0&0&0&0&0&0&0&0&0&0&0 \\
\hline
1&1&1&1&1&1&1&1&1&1&1&1 \\
\hline
2&2&2&2&2&2&2&2&2&2&2&2 \\
\hline
3&3&3&3&3&3&3&3&3&3&3&3 \\
\hline
4&4&4&4&4&4&4&4&4&4&4&4 \\
\hline
5&5&5&5&5&5&5&5&5&5&5&5 \\
\hline
6&6&6&6&6&6&6&6&6&6&6&6 \\
\hline
7&7&7&7&7&7&7&7&7&7&7&7 \\
\hline
8&8&8&8&8&8&8&8&8&8&8&8 \\
\hline
9&9&9&9&9&9&9&9&9&9&9&9 \\
\hline
10&10&10&10&10&10&10&10&10&10&10&10 \\
\hline
11&11&11&11&11&11&11&11&11&11&11&11 \\
\hline
\end{array}~~~~~
\begin{array}{|c|c|c|c|c|c|c|c|c|c|c|c|}
\hline
0&1&2&3&4&5&6&7&8&9&10&11 \\
\hline
1&2&3&4&5&0&7&8&9&10&11&6 \\
\hline
2&3&4&5&0&1&8&9&10&11&6&7 \\
\hline
3&4&5&0&1&2&9&10&11&6&7&8 \\
\hline
4&5&0&1&2&3&10&11&6&7&8&9 \\
\hline
5&0&1&2&3&4&11&6&7&8&9&10 \\
\hline
6&7&8&9&10&11&0&1&2&3&4&5 \\
\hline
7&8&9&10&11&6&1&2&3&4&5&0 \\
\hline
8&9&10&11&6&7&2&3&4&5&0&1 \\
\hline
9&10&11&6&7&8&3&4&5&0&1&2 \\
\hline
10&11&6&7&8&9&4&5&0&1&2&3 \\
\hline
11&6&7&8&9&10&5&0&1&2&3&4 \\
\hline
\end{array}~~~~~
\begin{array}{|c|c|c|c|c|c|c|c|c|c|c|c|}
\hline
0&3&6&1&9&11&2&8&5&4&7&10 \\
\hline
1&4&7&2&10&6&3&9&0&5&8&11 \\
\hline
2&5&8&3&11&7&4&10&1&0&9&6 \\
\hline
3&0&9&4&6&8&5&11&2&1&10&7 \\
\hline
4&1&10&5&7&9&0&6&3&2&11&8 \\
\hline
5&2&11&0&8&10&1&7&4&3&6&9 \\
\hline
6&9&0&7&3&5&8&2&11&10&1&4 \\
\hline
7&10&1&8&4&0&9&3&6&11&2&5 \\
\hline
8&11&2&9&5&1&10&4&7&6&3&0 \\
\hline
9&6&3&10&0&2&11&5&8&7&4&1 \\
\hline
10&7&4&11&1&3&6&0&9&8&5&2 \\
\hline
11&8&5&6&2&4&7&1&10&9&0&3 \\
\hline
\end{array}
\end{eqnarray*}
\vspace{-0.25cm}\hspace{3cm}$ L^0$\hspace{4.5cm}$L^1$\hspace{5cm}$L^2$
\begin{eqnarray*}
\footnotesize
 \setlength{\arraycolsep}{0.4 pt}
 \renewcommand\arraystretch{0.6}
\begin{array}{|c|c|c|c|c|c|c|c|c|c|c|c|}
\hline
0&8&1&11&5&9&3&10&2&7&6&4 \\
\hline
1&9&2&6&0&10&4&11&3&8&7&5 \\
\hline
2&10&3&7&1&11&5&6&4&9&8&0 \\
\hline
3&11&4&8&2&6&0&7&5&10&9&1 \\
\hline
4&6&5&9&3&7&1&8&6&11&10&2 \\
\hline
5&7&0&10&4&8&2&9&1&6&11&3 \\
\hline
6&2&7&5&11&3&9&4&8&1&0&10 \\
\hline
7&3&8&0&6&4&10&5&9&2&1&11 \\
\hline
8&4&9&1&7&5&11&0&10&3&2&6 \\
\hline
9&5&10&2&8&0&6&1&11&4&3&7 \\
\hline
10&0&11&3&9&1&7&2&6&5&4&8 \\
\hline
11&1&6&4&10&2&8&3&7&0&5&9 \\
\hline
\end{array}~~~~~
\begin{array}{|c|c|c|c|c|c|c|c|c|c|c|c|}
\hline
0&4&11&10&2&7&8&6&9&1&3&5 \\
\hline
1&5&6&11&3&8&9&7&10&2&4&0 \\
\hline
2&0&7&6&4&9&10&8&11&3&5&1 \\
\hline
3&1&8&7&5&10&11&9&6&4&0&2 \\
\hline
4&2&9&8&0&11&6&10&7&5&1&3 \\
\hline
5&3&10&9&1&6&7&11&8&0&2&4 \\
\hline
6&10&5&4&8&1&2&0&3&7&9&11 \\
\hline
7&11&0&5&9&2&3&1&4&8&10&6 \\
\hline
8&6&1&0&10&3&4&2&5&9&11&7 \\
\hline
9&7&2&1&11&4&5&3&0&10&6&8 \\
\hline
10&8&3&2&6&5&0&4&1&11&7&9 \\
\hline
11&9&4&3&7&0&1&5&2&6&8&10 \\
\hline
\end{array}~~~~~
\begin{array}{|c|c|c|c|c|c|c|c|c|c|c|c|}
\hline
0&6&8&2&7&1&9&11&4&10&5&3 \\
\hline
1&7&9&3&8&2&10&6&5&11&0&4 \\
\hline
2&8&10&4&9&3&11&7&0&6&1&5 \\
\hline
3&9&11&5&10&4&6&8&1&7&2&0 \\
\hline
4&10&6&0&11&5&7&9&2&8&3&1 \\
\hline
5&11&7&1&6&0&8&10&3&9&4&2 \\
\hline
6&0&2&8&1&7&3&5&10&4&11&9 \\
\hline
7&1&3&9&2&8&4&0&11&5&6&10 \\
\hline
8&2&4&10&3&9&5&1&6&0&7&11 \\
\hline
9&3&5&11&4&10&0&2&7&1&8&6 \\
\hline
10&4&0&6&5&11&1&3&8&2&9&7 \\
\hline
11&5&1&7&0&6&2&4&9&3&10&8 \\
\hline
\end{array}
\end{eqnarray*}
\vspace{-0.5cm}\hspace{3cm}$ L^3$\hspace{4.5cm}$L^4$\hspace{5cm}$L^5$

Actually, $L^i$ is a Latin square of order 12 on $\mathbb{Z}_{12}$ for $1\leq i\leq 5$. Set the complex Hadamard matrix $H_0=H_1=\ldots=H_5=\hat{F}_{12}$ with
\begin{eqnarray*}
\qihao
\begin{small}
\footnotesize
\hspace{-1.5cm} \setlength{\arraycolsep}{3 pt}
\renewcommand\arraystretch{0.3}
\hat{F}_{12}=\left(\begin{array}{cccccccccccc}
1&1&1&1&1&1&1&1&1&1&1&1 \\
1&\omega  &\omega^2&\omega^3&\omega^4&\omega^5&\omega^6&\omega^7&\omega^8&\omega^9&\omega^{10}&\omega^{11} \\
1&\omega^2  &\omega^4&\omega^6&\omega^8&\omega^{10}&    1&\omega^2&\omega^4&\omega^6&\omega^{8}&\omega^{10} \\
1&\omega^3  &\omega^6&\omega^9&       1&\omega^3&\omega^6&\omega^9&       1&\omega^3&\omega^{6}&\omega^{9} \\
1&\omega^4  &\omega^8&       1&\omega^4&\omega^8&     1&\omega^4&\omega^8&    1&\omega^{4}&\omega^{8} \\
1&\omega^5  &\omega^{10}&\omega^3&\omega^8&\omega&\omega^6&\omega^{11}&\omega^4&\omega^9&\omega^{2}&\omega^{7} \\
1&\omega^6  &       1&\omega^6&       1&\omega^6&    1 &\omega^6&    1&\omega^6&    1&\omega^{6} \\
1&\omega^7  &\omega^2&\omega^9&\omega^4&\omega^{11}&\omega^6&\omega^1&\omega^8&\omega^3&\omega^{10}&\omega^{5} \\
1&\omega^8  &\omega^4&       1&\omega^8&\omega^4&    1 &\omega^8&\omega^4&    1 &\omega^{8}&\omega^{4} \\
1&\omega^9  &\omega^6&\omega^3&       1&\omega^9&\omega^6&\omega^3&    1&\omega^9&\omega^{6}&\omega^{3} \\
1&\omega^{10}  &\omega^8&\omega^6&\omega^4&\omega^2&    1 &\omega^{10}&\omega^8&\omega^6&\omega^{4}&\omega^{2} \\
1&\omega^{11}  &\omega^{10}&\omega^9&\omega^8&\omega^7&\omega^6&\omega^5&\omega^4&\omega^3&\omega^{2}&\omega
\end{array}\right),
\end{small}
\end{eqnarray*}
where $\omega=e^{\frac{\pi \sqrt{-1}}{6}}$.
Then the six bases can be written as follows:
\begin{center}
\qihao
\begin{small}
 \setlength{\arraycolsep}{0.2 pt}
 \renewcommand\arraystretch{0.59}
  $  \left\{
          \begin{array}{ll}
A^0_{0,j}=\frac{1}{2\sqrt{3}}|0\rangle (|0\rangle+\omega^j|1\rangle+\omega^{2j}|2\rangle+\omega^{3j}|3\rangle+\omega^{4j}|4\rangle+\omega^{5j}|5\rangle+\omega^{6j}|6\rangle
+\omega^{7j}|7\rangle+\omega^{8j}|8\rangle\\
\hspace{1.1cm}+\omega^{9j}|9\rangle+\omega^{10j}|10\rangle+\omega^{11j}|11\rangle) \\
A^0_{1,j}=\frac{1}{2\sqrt{3}}|1\rangle (|0\rangle+\omega^j|1\rangle+\omega^{2j}|2\rangle+\omega^{3j}|3\rangle+\omega^{4j}|4\rangle+\omega^{5j}|5\rangle+\omega^{6j}|6\rangle
+\omega^{7j}|7\rangle+\omega^{8j}|8\rangle\\
\hspace{1.1cm}+\omega^{9j}|9\rangle+\omega^{10j}|10\rangle+\omega^{11j}|11\rangle) \\
\hspace{1cm}\cdots\\
A^0_{11,j}=\frac{1}{2\sqrt{3}}|11\rangle (|0\rangle+\omega^j|1\rangle+\omega^{2j}|2\rangle+\omega^{3j}|3\rangle+\omega^{4j}|4\rangle+\omega^{5j}|5\rangle+\omega^{6j}|6\rangle
+\omega^{7j}|7\rangle+\omega^{8j}|8\rangle\\
\hspace{1.1cm}+\omega^{9j}|9\rangle+\omega^{10j}|10\rangle+\omega^{11j}|11\rangle)
     \end{array}
       \right.$
\end{small}
\end{center}\newpage
\begin{center}
\footnotesize
 \setlength{\arraycolsep}{0.0 pt}
 \renewcommand\arraystretch{0.59}
  $  \left\{
          \begin{array}{ll}
A^1_{0,j}=\frac{1}{2\sqrt{3}}(|0\rangle |0\rangle +\omega^j|1\rangle|1\rangle+\omega^{2j}|2\rangle|2\rangle+\omega^{3j}|3\rangle|3\rangle+\omega^{4j}|4\rangle|4\rangle+\omega^{5j}|5\rangle|5\rangle
+\omega^{6j}|6\rangle|6\rangle\\
\hspace{1.1cm} +\omega^{7j}|7\rangle|7\rangle+\omega^{8j}|8\rangle|8\rangle
+\omega^{9j}|9\rangle|9\rangle+\omega^{10j}|10\rangle|10\rangle+\omega^{11j}|11\rangle|11\rangle) \\
A^1_{1,j}=\frac{1}{2\sqrt{3}} (|1\rangle|0\rangle+\omega^j|2\rangle|1\rangle+\omega^{2j}|3\rangle|2\rangle+\omega^{3j}|4\rangle|3\rangle+\omega^{4j}|5\rangle|4\rangle+
\omega^{5j}|0\rangle|5\rangle+\omega^{6j}|7\rangle|6\rangle\\
\hspace{1.1cm}
+\omega^{7j}|8\rangle|7\rangle+\omega^{8j}|9\rangle|8\rangle+\omega^{9j}|10\rangle|9\rangle+\omega^{10j}|11\rangle|10\rangle+\omega^{11j}|6\rangle|11\rangle) \\
A^1_{2,j}=\frac{1}{2\sqrt{3}}(|2\rangle |0\rangle +\omega^j|3\rangle|1\rangle+\omega^{2j}|4\rangle|2\rangle+\omega^{3j}|5\rangle|3\rangle+\omega^{4j}|0\rangle|4\rangle+\omega^{5j}|1\rangle|5\rangle
+\omega^{6j}|8\rangle|6\rangle\\
\hspace{1.1cm} +\omega^{7j}|9\rangle|7\rangle+\omega^{8j}|10\rangle|8\rangle
+\omega^{9j}|11\rangle|9\rangle+\omega^{10j}|6\rangle|10\rangle+\omega^{11j}|7\rangle|11\rangle) \\
A^1_{3,j}=\frac{1}{2\sqrt{3}}(|3\rangle |0\rangle +\omega^j|4\rangle|1\rangle+\omega^{2j}|5\rangle|2\rangle+\omega^{3j}|0\rangle|3\rangle+\omega^{4j}|1\rangle|4\rangle+\omega^{5j}|2\rangle|5\rangle
+\omega^{6j}|9\rangle|6\rangle\\
\hspace{1.1cm} +\omega^{7j}|10\rangle|7\rangle+\omega^{8j}|11\rangle|8\rangle
+\omega^{9j}|6\rangle|9\rangle+\omega^{10j}|7\rangle|10\rangle+\omega^{11j}|8\rangle|11\rangle) \\
A^1_{4,j}=\frac{1}{2\sqrt{3}}(|4\rangle |0\rangle +\omega^j|5\rangle|1\rangle+\omega^{2j}|0\rangle|2\rangle+\omega^{3j}|1\rangle|3\rangle+\omega^{4j}|2\rangle|4\rangle+\omega^{5j}|3\rangle|5\rangle
+\omega^{6j}|10\rangle|6\rangle\\
\hspace{1.1cm} +\omega^{7j}|11\rangle|7\rangle+\omega^{8j}|6\rangle|8\rangle
+\omega^{9j}|7\rangle|9\rangle+\omega^{10j}|8\rangle|10\rangle+\omega^{11j}|9\rangle|11\rangle) \\
A^1_{5,j}=\frac{1}{2\sqrt{3}}(|5\rangle |0\rangle +\omega^j|0\rangle|1\rangle+\omega^{2j}|1\rangle|2\rangle+\omega^{3j}|2\rangle|3\rangle+\omega^{4j}|3\rangle|4\rangle+\omega^{5j}|4\rangle|5\rangle
+\omega^{6j}|11\rangle|6\rangle\\
\hspace{1.1cm} +\omega^{7j}|6\rangle|7\rangle+\omega^{8j}|7\rangle|8\rangle
+\omega^{9j}|8\rangle|9\rangle+\omega^{10j}|9\rangle|10\rangle+\omega^{11j}|10\rangle|11\rangle) \\
A^1_{6,j}=\frac{1}{2\sqrt{3}}(|6\rangle |0\rangle +\omega^j|7\rangle|1\rangle+\omega^{2j}|8\rangle|2\rangle+\omega^{3j}|9\rangle|3\rangle+\omega^{4j}|10\rangle|4\rangle+\omega^{5j}|11\rangle|5\rangle
+\omega^{6j}|0\rangle|6\rangle\\
\hspace{1.1cm} +\omega^{7j}|1\rangle|7\rangle+\omega^{8j}|2\rangle|8\rangle
+\omega^{9j}|3\rangle|9\rangle+\omega^{10j}|4\rangle|10\rangle+\omega^{11j}|5\rangle|11\rangle) \\
A^1_{7,j}=\frac{1}{2\sqrt{3}}(|7\rangle |0\rangle +\omega^j|8\rangle|1\rangle+\omega^{2j}|9\rangle|2\rangle+\omega^{3j}|10\rangle|3\rangle+\omega^{4j}|11\rangle|4\rangle+\omega^{5j}|6\rangle|5\rangle
+\omega^{6j}|1\rangle|6\rangle\\
\hspace{1.1cm} +\omega^{7j}|2\rangle|7\rangle+\omega^{8j}|3\rangle|8\rangle
+\omega^{9j}|4\rangle|9\rangle+\omega^{10j}|5\rangle|10\rangle+\omega^{11j}|0\rangle|11\rangle) \\
A^1_{8,j}=\frac{1}{2\sqrt{3}}(|8\rangle |0\rangle +\omega^j|9\rangle|1\rangle+\omega^{2j}|10\rangle|2\rangle+\omega^{3j}|11\rangle|3\rangle+\omega^{4j}|6\rangle|4\rangle+\omega^{5j}|7\rangle|5\rangle
+\omega^{6j}|2\rangle|6\rangle\\
\hspace{1.1cm} +\omega^{7j}|3\rangle|7\rangle+\omega^{8j}|4\rangle|8\rangle
+\omega^{9j}|5\rangle|9\rangle+\omega^{10j}|0\rangle|10\rangle+\omega^{11j}|1\rangle|11\rangle) \\
A^1_{9,j}=\frac{1}{2\sqrt{3}}(|9\rangle |0\rangle +\omega^j|10\rangle|1\rangle+\omega^{2j}|11\rangle|2\rangle+\omega^{3j}|6\rangle|3\rangle+\omega^{4j}|7\rangle|4\rangle+\omega^{5j}|8\rangle|5\rangle
+\omega^{6j}|3\rangle|6\rangle\\
\hspace{1.1cm} +\omega^{7j}|4\rangle|7\rangle+\omega^{8j}|5\rangle|8\rangle
+\omega^{9j}|0\rangle|9\rangle+\omega^{10j}|1\rangle|10\rangle+\omega^{11j}|2\rangle|11\rangle) \\
A^1_{10,j}=\frac{1}{2\sqrt{3}}(|10\rangle |0\rangle +\omega^j|11\rangle|1\rangle+\omega^{2j}|6\rangle|2\rangle+\omega^{3j}|7\rangle|3\rangle+\omega^{4j}|8\rangle|4\rangle+\omega^{5j}|9\rangle|5\rangle
+\omega^{6j}|4\rangle|6\rangle\\
\hspace{1.1cm} +\omega^{7j}|5\rangle|7\rangle+\omega^{8j}|0\rangle|8\rangle
+\omega^{9j}|1\rangle|9\rangle+\omega^{10j}|2\rangle|10\rangle+\omega^{11j}|3\rangle|11\rangle) \\
A^1_{11,j}=\frac{1}{2\sqrt{3}}(|11\rangle |0\rangle +\omega^j|6\rangle|1\rangle+\omega^{2j}|7\rangle|2\rangle+\omega^{3j}|8\rangle|3\rangle+\omega^{4j}|9\rangle|4\rangle+\omega^{5j}|10\rangle|5\rangle
+\omega^{6j}|5\rangle|6\rangle\\
\hspace{1.1cm} +\omega^{7j}|0\rangle|7\rangle+\omega^{8j}|1\rangle|8\rangle
+\omega^{9j}|2\rangle|9\rangle+\omega^{10j}|3\rangle|10\rangle+\omega^{11j}|4\rangle|11\rangle) \\
     \end{array}
       \right.$
\end{center}
\begin{center}
\footnotesize
 \setlength{\arraycolsep}{0.0 pt}
 \renewcommand\arraystretch{0.59}
 $  \left\{
       \begin{array}{ll}
A^2_{0,j}=\frac{1}{2\sqrt{3}}(|0\rangle |0\rangle +\omega^j|3\rangle|1\rangle+\omega^{2j}|6\rangle|2\rangle+\omega^{3j}|1\rangle|3\rangle+\omega^{4j}|9\rangle|4\rangle+\omega^{5j}|11\rangle|5\rangle
+\omega^{6j}|2\rangle|6\rangle\\
\hspace{1.1cm} +\omega^{7j}|8\rangle|7\rangle+\omega^{8j}|5\rangle|8\rangle
+\omega^{9j}|4\rangle|9\rangle+\omega^{10j}|7\rangle|10\rangle+\omega^{11j}|10\rangle|11\rangle) \\
A^2_{1,j}=\frac{1}{2\sqrt{3}} (|1\rangle|1\rangle+\omega^j|4\rangle|1\rangle+\omega^{2j}|7\rangle|2\rangle+\omega^{3j}|2\rangle|3\rangle+\omega^{4j}|10\rangle|4\rangle+
\omega^{5j}|6\rangle|5\rangle+\omega^{6j}|3\rangle|6\rangle\\
\hspace{1.1cm}
+\omega^{7j}|9\rangle|7\rangle+\omega^{8j}|0\rangle|8\rangle+\omega^{9j}|5\rangle|9\rangle+\omega^{10j}|8\rangle|10\rangle+\omega^{11j}|11\rangle|11\rangle) \\
A^2_{2,j}=\frac{1}{2\sqrt{3}}(|2\rangle |0\rangle +\omega^j|5\rangle|1\rangle+\omega^{2j}|8\rangle|2\rangle+\omega^{3j}|3\rangle|3\rangle+\omega^{4j}|11\rangle|4\rangle+\omega^{5j}|7\rangle|5\rangle
+\omega^{6j}|4\rangle|6\rangle\\
\hspace{1.1cm} +\omega^{7j}|10\rangle|7\rangle+\omega^{8j}|1\rangle|8\rangle
+\omega^{9j}|0\rangle|9\rangle+\omega^{10j}|9\rangle|10\rangle+\omega^{11j}|6\rangle|11\rangle) \\
A^2_{3,j}=\frac{1}{2\sqrt{3}}(|3\rangle |0\rangle +\omega^j|0\rangle|1\rangle+\omega^{2j}|9\rangle|2\rangle+\omega^{3j}|4\rangle|3\rangle+\omega^{4j}|6\rangle|4\rangle+\omega^{5j}|8\rangle|5\rangle
+\omega^{6j}|5\rangle|6\rangle\\
\hspace{1.1cm} +\omega^{7j}|11\rangle|7\rangle+\omega^{8j}|2\rangle|8\rangle
+\omega^{9j}|1\rangle|9\rangle+\omega^{10j}|10\rangle|10\rangle+\omega^{11j}|7\rangle|11\rangle) \\
A^2_{4,j}=\frac{1}{2\sqrt{3}}(|4\rangle |0\rangle +\omega^j|1\rangle|1\rangle+\omega^{2j}|10\rangle|2\rangle+\omega^{3j}|5\rangle|3\rangle+\omega^{4j}|7\rangle|4\rangle+\omega^{5j}|9\rangle|5\rangle
+\omega^{6j}|0\rangle|6\rangle\\
\hspace{1.1cm} +\omega^{7j}|6\rangle|7\rangle+\omega^{8j}|3\rangle|8\rangle
+\omega^{9j}|2\rangle|9\rangle+\omega^{10j}|11\rangle|10\rangle+\omega^{11j}|8\rangle|11\rangle) \\
A^2_{5,j}=\frac{1}{2\sqrt{3}}(|5\rangle |0\rangle +\omega^j|2\rangle|1\rangle+\omega^{2j}|11\rangle|2\rangle+\omega^{3j}|0\rangle|3\rangle+\omega^{4j}|8\rangle|4\rangle+\omega^{5j}|10\rangle|5\rangle
+\omega^{6j}|1\rangle|6\rangle\\
\hspace{1.1cm} +\omega^{7j}|7\rangle|7\rangle+\omega^{8j}|4\rangle|8\rangle
+\omega^{9j}|3\rangle|9\rangle+\omega^{10j}|6\rangle|10\rangle+\omega^{11j}|9\rangle|11\rangle) \\
A^2_{6,j}=\frac{1}{2\sqrt{3}}(|6\rangle |0\rangle +\omega^j|9\rangle|1\rangle+\omega^{2j}|0\rangle|2\rangle+\omega^{3j}|7\rangle|3\rangle+\omega^{4j}|3\rangle|4\rangle+\omega^{5j}|5\rangle|5\rangle
+\omega^{6j}|8\rangle|6\rangle\\
\hspace{1.1cm} +\omega^{7j}|2\rangle|7\rangle+\omega^{8j}|11\rangle|8\rangle
+\omega^{9j}|10\rangle|9\rangle+\omega^{10j}|1\rangle|10\rangle+\omega^{11j}|4\rangle|11\rangle) \\
A^2_{7,j}=\frac{1}{2\sqrt{3}}(|7\rangle |0\rangle +\omega^j|10\rangle|1\rangle+\omega^{2j}|1\rangle|2\rangle+\omega^{3j}|8\rangle|3\rangle+\omega^{4j}|4\rangle|4\rangle+\omega^{5j}|0\rangle|5\rangle
+\omega^{6j}|9\rangle|6\rangle\\
\hspace{1.1cm} +\omega^{7j}|3\rangle|7\rangle+\omega^{8j}|6\rangle|8\rangle
+\omega^{9j}|11\rangle|9\rangle+\omega^{10j}|2\rangle|10\rangle+\omega^{11j}|5\rangle|11\rangle) \\
A^2_{8,j}=\frac{1}{2\sqrt{3}}(|8\rangle |0\rangle +\omega^j|11\rangle|1\rangle+\omega^{2j}|2\rangle|2\rangle+\omega^{3j}|9\rangle|3\rangle+\omega^{4j}|5\rangle|4\rangle+\omega^{5j}|1\rangle|5\rangle
+\omega^{6j}|10\rangle|6\rangle\\
\hspace{1.1cm} +\omega^{7j}|4\rangle|7\rangle+\omega^{8j}|7\rangle|8\rangle
+\omega^{9j}|6\rangle|9\rangle+\omega^{10j}|3\rangle|10\rangle+\omega^{11j}|0\rangle|11\rangle) \\
A^2_{9,j}=\frac{1}{2\sqrt{3}}(|9\rangle |0\rangle +\omega^j|6\rangle|1\rangle+\omega^{2j}|3\rangle|2\rangle+\omega^{3j}|10\rangle|3\rangle+\omega^{4j}|0\rangle|4\rangle+\omega^{5j}|2\rangle|5\rangle
+\omega^{6j}|11\rangle|6\rangle\\
\hspace{1.1cm} +\omega^{7j}|5\rangle|7\rangle+\omega^{8j}|8\rangle|8\rangle
+\omega^{9j}|7\rangle|9\rangle+\omega^{10j}|4\rangle|10\rangle+\omega^{11j}|1\rangle|11\rangle) \\
A^2_{10,j}=\frac{1}{2\sqrt{3}}(|10\rangle |0\rangle +\omega^j|7\rangle|1\rangle+\omega^{2j}|4\rangle|2\rangle+\omega^{3j}|11\rangle|3\rangle+\omega^{4j}|1\rangle|4\rangle+\omega^{5j}|3\rangle|5\rangle
+\omega^{6j}|6\rangle|6\rangle\\
\hspace{1.1cm} +\omega^{7j}|0\rangle|7\rangle+\omega^{8j}|9\rangle|8\rangle
+\omega^{9j}|8\rangle|9\rangle+\omega^{10j}|5\rangle|10\rangle+\omega^{11j}|2\rangle|11\rangle) \\
A^2_{11,j}=\frac{1}{2\sqrt{3}}(|11\rangle |0\rangle +\omega^j|8\rangle|1\rangle+\omega^{2j}|5\rangle|2\rangle+\omega^{3j}|6\rangle|3\rangle+\omega^{4j}|2\rangle|4\rangle+\omega^{5j}|4\rangle|5\rangle
+\omega^{6j}|7\rangle|6\rangle\\
\hspace{1.1cm} +\omega^{7j}|1\rangle|7\rangle+\omega^{8j}|10\rangle|8\rangle
+\omega^{9j}|9\rangle|9\rangle+\omega^{10j}|0\rangle|10\rangle+\omega^{11j}|3\rangle|11\rangle) \\
     \end{array}
       \right.$
\end{center}
\begin{center}
\footnotesize
 \setlength{\arraycolsep}{0.0 pt}
 \renewcommand\arraystretch{0.59}
  $  \left\{
          \begin{array}{ll}
A^3_{0,j}=\frac{1}{2\sqrt{3}}(|0\rangle |0\rangle +\omega^j|8\rangle|1\rangle+\omega^{2j}|1\rangle|2\rangle+\omega^{3j}|11\rangle|3\rangle+\omega^{4j}|5\rangle|4\rangle+\omega^{5j}|9\rangle|5\rangle
+\omega^{6j}|3\rangle|6\rangle\\
\hspace{1.1cm} +\omega^{7j}|10\rangle|7\rangle+\omega^{8j}|2\rangle|8\rangle
+\omega^{9j}|7\rangle|9\rangle+\omega^{10j}|6\rangle|10\rangle+\omega^{11j}|4\rangle|11\rangle) \\
A^3_{1,j}=\frac{1}{2\sqrt{3}} (|1\rangle|1\rangle+\omega^j|9\rangle|1\rangle+\omega^{2j}|2\rangle|2\rangle+\omega^{3j}|6\rangle|3\rangle+\omega^{4j}|0\rangle|4\rangle+
\omega^{5j}|10\rangle|5\rangle+\omega^{6j}|4\rangle|6\rangle\\
\hspace{1.1cm}
+\omega^{7j}|11\rangle|7\rangle+\omega^{8j}|3\rangle|8\rangle+\omega^{9j}|8\rangle|9\rangle+\omega^{10j}|7\rangle|10\rangle+\omega^{11j}|5\rangle|11\rangle) \\
A^3_{2,j}=\frac{1}{2\sqrt{3}}(|2\rangle |0\rangle +\omega^j|10\rangle|1\rangle+\omega^{2j}|3\rangle|2\rangle+\omega^{3j}|7\rangle|3\rangle+\omega^{4j}|1\rangle|4\rangle+\omega^{5j}|11\rangle|5\rangle
+\omega^{6j}|5\rangle|6\rangle\\
\hspace{1.1cm} +\omega^{7j}|6\rangle|7\rangle+\omega^{8j}|4\rangle|8\rangle
+\omega^{9j}|9\rangle|9\rangle+\omega^{10j}|8\rangle|10\rangle+\omega^{11j}|0\rangle|11\rangle) \\
A^3_{3,j}=\frac{1}{2\sqrt{3}}(|3\rangle |0\rangle +\omega^j|11\rangle|1\rangle+\omega^{2j}|4\rangle|2\rangle+\omega^{3j}|8\rangle|3\rangle+\omega^{4j}|2\rangle|4\rangle+\omega^{5j}|6\rangle|5\rangle
+\omega^{6j}|0\rangle|6\rangle\\
\hspace{1.1cm} +\omega^{7j}|7\rangle|7\rangle+\omega^{8j}|5\rangle|8\rangle
+\omega^{9j}|10\rangle|9\rangle+\omega^{10j}|9\rangle|10\rangle+\omega^{11j}|1\rangle|11\rangle) \\
A^3_{4,j}=\frac{1}{2\sqrt{3}}(|4\rangle |0\rangle +\omega^j|6\rangle|1\rangle+\omega^{2j}|5\rangle|2\rangle+\omega^{3j}|9\rangle|3\rangle+\omega^{4j}|3\rangle|4\rangle+\omega^{5j}|7\rangle|5\rangle
+\omega^{6j}|1\rangle|6\rangle\\
\hspace{1.1cm} +\omega^{7j}|8\rangle|7\rangle+\omega^{8j}|6\rangle|8\rangle
+\omega^{9j}|11\rangle|9\rangle+\omega^{10j}|10\rangle|10\rangle+\omega^{11j}|2\rangle|11\rangle) \\
A^3_{5,j}=\frac{1}{2\sqrt{3}}(|5\rangle |0\rangle +\omega^j|7\rangle|1\rangle+\omega^{2j}|0\rangle|2\rangle+\omega^{3j}|10\rangle|3\rangle+\omega^{4j}|4\rangle|4\rangle+\omega^{5j}|8\rangle|5\rangle
+\omega^{6j}|2\rangle|6\rangle\\
\hspace{1.1cm} +\omega^{7j}|9\rangle|7\rangle+\omega^{8j}|1\rangle|8\rangle
+\omega^{9j}|6\rangle|9\rangle+\omega^{10j}|11\rangle|10\rangle+\omega^{11j}|3\rangle|11\rangle) \\
A^3_{6,j}=\frac{1}{2\sqrt{3}}(|6\rangle |0\rangle +\omega^j|2\rangle|1\rangle+\omega^{2j}|7\rangle|2\rangle+\omega^{3j}|5\rangle|3\rangle+\omega^{4j}|11\rangle|4\rangle+\omega^{5j}|3\rangle|5\rangle
+\omega^{6j}|9\rangle|6\rangle\\
\hspace{1.1cm} +\omega^{7j}|4\rangle|7\rangle+\omega^{8j}|8\rangle|8\rangle
+\omega^{9j}|1\rangle|9\rangle+\omega^{10j}|0\rangle|10\rangle+\omega^{11j}|10\rangle|11\rangle) \\
A^3_{7,j}=\frac{1}{2\sqrt{3}}(|7\rangle |0\rangle +\omega^j|3\rangle|1\rangle+\omega^{2j}|8\rangle|2\rangle+\omega^{3j}|0\rangle|3\rangle+\omega^{4j}|6\rangle|4\rangle+\omega^{5j}|4\rangle|5\rangle
+\omega^{6j}|10\rangle|6\rangle\\
\hspace{1.1cm} +\omega^{7j}|5\rangle|7\rangle+\omega^{8j}|9\rangle|8\rangle
+\omega^{9j}|2\rangle|9\rangle+\omega^{10j}|1\rangle|10\rangle+\omega^{11j}|11\rangle|11\rangle) \\
A^3_{8,j}=\frac{1}{2\sqrt{3}}(|8\rangle |0\rangle +\omega^j|4\rangle|1\rangle+\omega^{2j}|9\rangle|2\rangle+\omega^{3j}|1\rangle|3\rangle+\omega^{4j}|7\rangle|4\rangle+\omega^{5j}|5\rangle|5\rangle
+\omega^{6j}|11\rangle|6\rangle\\
\hspace{1.1cm} +\omega^{7j}|0\rangle|7\rangle+\omega^{8j}|10\rangle|8\rangle
+\omega^{9j}|3\rangle|9\rangle+\omega^{10j}|2\rangle|10\rangle+\omega^{11j}|6\rangle|11\rangle) \\
A^3_{9,j}=\frac{1}{2\sqrt{3}}(|9\rangle |0\rangle +\omega^j|5\rangle|1\rangle+\omega^{2j}|10\rangle|2\rangle+\omega^{3j}|2\rangle|3\rangle+\omega^{4j}|8\rangle|4\rangle+\omega^{5j}|0\rangle|5\rangle
+\omega^{6j}|6\rangle|6\rangle\\
\hspace{1.1cm} +\omega^{7j}|1\rangle|7\rangle+\omega^{8j}|11\rangle|8\rangle
+\omega^{9j}|4\rangle|9\rangle+\omega^{10j}|3\rangle|10\rangle+\omega^{11j}|7\rangle|11\rangle) \\
A^3_{10,j}=\frac{1}{2\sqrt{3}}(|10\rangle |0\rangle +\omega^j|0\rangle|1\rangle+\omega^{2j}|11\rangle|2\rangle+\omega^{3j}|3\rangle|3\rangle+\omega^{4j}|9\rangle|4\rangle+\omega^{5j}|1\rangle|5\rangle
+\omega^{6j}|7\rangle|6\rangle\\
\hspace{1.1cm} +\omega^{7j}|2\rangle|7\rangle+\omega^{8j}|6\rangle|8\rangle
+\omega^{9j}|5\rangle|9\rangle+\omega^{10j}|4\rangle|10\rangle+\omega^{11j}|8\rangle|11\rangle) \\
A^3_{11,j}=\frac{1}{2\sqrt{3}}(|11\rangle |0\rangle +\omega^j|1\rangle|1\rangle+\omega^{2j}|6\rangle|2\rangle+\omega^{3j}|4\rangle|3\rangle+\omega^{4j}|10\rangle|4\rangle+\omega^{5j}|2\rangle|5\rangle
+\omega^{6j}|8\rangle|6\rangle\\
\hspace{1.1cm} +\omega^{7j}|3\rangle|7\rangle+\omega^{8j}|7\rangle|8\rangle
+\omega^{9j}|0\rangle|9\rangle+\omega^{10j}|5\rangle|10\rangle+\omega^{11j}|9\rangle|11\rangle) \\
     \end{array}
       \right.$
\end{center}
\begin{center}
\footnotesize
 \setlength{\arraycolsep}{0.0 pt}
 \renewcommand\arraystretch{0.59}
  $  \left\{
          \begin{array}{ll}
A^4_{0,j}=\frac{1}{2\sqrt{3}}(|0\rangle |0\rangle +\omega^j|4\rangle|1\rangle+\omega^{2j}|11\rangle|2\rangle+\omega^{3j}|10\rangle|3\rangle+\omega^{4j}|2\rangle|4\rangle+\omega^{5j}|7\rangle|5\rangle
+\omega^{6j}|8\rangle|6\rangle\\
\hspace{1.1cm} +\omega^{7j}|6\rangle|7\rangle+\omega^{8j}|9\rangle|8\rangle
+\omega^{9j}|1\rangle|9\rangle+\omega^{10j}|3\rangle|10\rangle+\omega^{11j}|5\rangle|11\rangle) \\
A^4_{1,j}=\frac{1}{2\sqrt{3}} (|1\rangle|1\rangle+\omega^j|5\rangle|1\rangle+\omega^{2j}|6\rangle|2\rangle+\omega^{3j}|11\rangle|3\rangle+\omega^{4j}|3\rangle|4\rangle+
\omega^{5j}|8\rangle|5\rangle+\omega^{6j}|9\rangle|6\rangle\\
\hspace{1.1cm}
+\omega^{7j}|7\rangle|7\rangle+\omega^{8j}|10\rangle|8\rangle+\omega^{9j}|2\rangle|9\rangle+\omega^{10j}|4\rangle|10\rangle+\omega^{11j}|0\rangle|11\rangle) \\
A^4_{2,j}=\frac{1}{2\sqrt{3}}(|2\rangle |0\rangle +\omega^j|0\rangle|1\rangle+\omega^{2j}|7\rangle|2\rangle+\omega^{3j}|6\rangle|3\rangle+\omega^{4j}|4\rangle|4\rangle+\omega^{5j}|9\rangle|5\rangle
+\omega^{6j}|10\rangle|6\rangle\\
\hspace{1.1cm} +\omega^{7j}|8\rangle|7\rangle+\omega^{8j}|11\rangle|8\rangle
+\omega^{9j}|3\rangle|9\rangle+\omega^{10j}|5\rangle|10\rangle+\omega^{11j}|1\rangle|11\rangle) \\
A^4_{3,j}=\frac{1}{2\sqrt{3}}(|3\rangle |0\rangle +\omega^j|1\rangle|1\rangle+\omega^{2j}|8\rangle|2\rangle+\omega^{3j}|7\rangle|3\rangle+\omega^{4j}|5\rangle|4\rangle+\omega^{5j}|10\rangle|5\rangle
+\omega^{6j}|11\rangle|6\rangle\\
\hspace{1.1cm} +\omega^{7j}|9\rangle|7\rangle+\omega^{8j}|6\rangle|8\rangle
+\omega^{9j}|4\rangle|9\rangle+\omega^{10j}|0\rangle|10\rangle+\omega^{11j}|2\rangle|11\rangle) \\
A^4_{4,j}=\frac{1}{2\sqrt{3}}(|4\rangle |0\rangle +\omega^j|2\rangle|1\rangle+\omega^{2j}|9\rangle|2\rangle+\omega^{3j}|8\rangle|3\rangle+\omega^{4j}|0\rangle|4\rangle+\omega^{5j}|11\rangle|5\rangle
+\omega^{6j}|6\rangle|6\rangle\\
\hspace{1.1cm} +\omega^{7j}|10\rangle|7\rangle+\omega^{8j}|7\rangle|8\rangle
+\omega^{9j}|5\rangle|9\rangle+\omega^{10j}|1\rangle|10\rangle+\omega^{11j}|3\rangle|11\rangle) \\
A^4_{5,j}=\frac{1}{2\sqrt{3}}(|5\rangle |0\rangle +\omega^j|3\rangle|1\rangle+\omega^{2j}|10\rangle|2\rangle+\omega^{3j}|9\rangle|3\rangle+\omega^{4j}|1\rangle|4\rangle+\omega^{5j}|6\rangle|5\rangle
+\omega^{6j}|7\rangle|6\rangle\\
\hspace{1.1cm} +\omega^{7j}|11\rangle|7\rangle+\omega^{8j}|8\rangle|8\rangle
+\omega^{9j}|0\rangle|9\rangle+\omega^{10j}|2\rangle|10\rangle+\omega^{11j}|4\rangle|11\rangle) \\
A^4_{6,j}=\frac{1}{2\sqrt{3}}(|6\rangle |0\rangle +\omega^j|10\rangle|1\rangle+\omega^{2j}|5\rangle|2\rangle+\omega^{3j}|4\rangle|3\rangle+\omega^{4j}|8\rangle|4\rangle+\omega^{5j}|1\rangle|5\rangle
+\omega^{6j}|2\rangle|6\rangle\\
\hspace{1.1cm} +\omega^{7j}|0\rangle|7\rangle+\omega^{8j}|3\rangle|8\rangle
+\omega^{9j}|7\rangle|9\rangle+\omega^{10j}|9\rangle|10\rangle+\omega^{11j}|11\rangle|11\rangle) \\
A^4_{7,j}=\frac{1}{2\sqrt{3}}(|7\rangle |0\rangle +\omega^j|11\rangle|1\rangle+\omega^{2j}|0\rangle|2\rangle+\omega^{3j}|5\rangle|3\rangle+\omega^{4j}|9\rangle|4\rangle+\omega^{5j}|2\rangle|5\rangle
+\omega^{6j}|3\rangle|6\rangle\\
\hspace{1.1cm} +\omega^{7j}|1\rangle|7\rangle+\omega^{8j}|4\rangle|8\rangle
+\omega^{9j}|8\rangle|9\rangle+\omega^{10j}|10\rangle|10\rangle+\omega^{11j}|6\rangle|11\rangle) \\
A^4_{8,j}=\frac{1}{2\sqrt{3}}(|8\rangle |0\rangle +\omega^j|6\rangle|1\rangle+\omega^{2j}|1\rangle|2\rangle+\omega^{3j}|0\rangle|3\rangle+\omega^{4j}|10\rangle|4\rangle+\omega^{5j}|3\rangle|5\rangle
+\omega^{6j}|4\rangle|6\rangle\\
\hspace{1.1cm} +\omega^{7j}|2\rangle|7\rangle+\omega^{8j}|5\rangle|8\rangle
+\omega^{9j}|9\rangle|9\rangle+\omega^{10j}|11\rangle|10\rangle+\omega^{11j}|7\rangle|11\rangle) \\
A^4_{9,j}=\frac{1}{2\sqrt{3}}(|9\rangle |0\rangle +\omega^j|7\rangle|1\rangle+\omega^{2j}|2\rangle|2\rangle+\omega^{3j}|1\rangle|3\rangle+\omega^{4j}|11\rangle|4\rangle+\omega^{5j}|4\rangle|5\rangle
+\omega^{6j}|5\rangle|6\rangle\\
\hspace{1.1cm} +\omega^{7j}|3\rangle|7\rangle+\omega^{8j}|0\rangle|8\rangle
+\omega^{9j}|10\rangle|9\rangle+\omega^{10j}|6\rangle|10\rangle+\omega^{11j}|8\rangle|11\rangle) \\
A^4_{10,j}=\frac{1}{2\sqrt{3}}(|10\rangle |0\rangle +\omega^j|8\rangle|1\rangle+\omega^{2j}|3\rangle|2\rangle+\omega^{3j}|2\rangle|3\rangle+\omega^{4j}|6\rangle|4\rangle+\omega^{5j}|5\rangle|5\rangle
+\omega^{6j}|0\rangle|6\rangle\\
\hspace{1.1cm} +\omega^{7j}|4\rangle|7\rangle+\omega^{8j}|1\rangle|8\rangle
+\omega^{9j}|11\rangle|9\rangle+\omega^{10j}|7\rangle|10\rangle+\omega^{11j}|9\rangle|11\rangle) \\
A^4_{11,j}=\frac{1}{2\sqrt{3}}(|11\rangle |0\rangle +\omega^j|9\rangle|1\rangle+\omega^{2j}|4\rangle|2\rangle+\omega^{3j}|3\rangle|3\rangle+\omega^{4j}|7\rangle|4\rangle+\omega^{5j}|0\rangle|5\rangle
+\omega^{6j}|1\rangle|6\rangle\\
\hspace{1.1cm} +\omega^{7j}|5\rangle|7\rangle+\omega^{8j}|2\rangle|8\rangle
+\omega^{9j}|6\rangle|9\rangle+\omega^{10j}|8\rangle|10\rangle+\omega^{11j}|10\rangle|11\rangle) \\
     \end{array}
       \right.$
\end{center}
\begin{center}
\footnotesize
 \setlength{\arraycolsep}{0.0 pt}
 \renewcommand\arraystretch{0.59}
  $  \left\{
          \begin{array}{ll}
A^5_{0,j}=\frac{1}{2\sqrt{3}}(|0\rangle |0\rangle +\omega^j|6\rangle|1\rangle+\omega^{2j}|8\rangle|2\rangle+\omega^{3j}|2\rangle|3\rangle+\omega^{4j}|7\rangle|4\rangle+\omega^{5j}|1\rangle|5\rangle
+\omega^{6j}|9\rangle|6\rangle\\
\hspace{1.1cm} +\omega^{7j}|11\rangle|7\rangle+\omega^{8j}|4\rangle|8\rangle
+\omega^{9j}|10\rangle|9\rangle+\omega^{10j}|5\rangle|10\rangle+\omega^{11j}|3\rangle|11\rangle) \\
A^5_{1,j}=\frac{1}{2\sqrt{3}} (|1\rangle|1\rangle+\omega^j|7\rangle|1\rangle+\omega^{2j}|9\rangle|2\rangle+\omega^{3j}|3\rangle|3\rangle+\omega^{4j}|8\rangle|4\rangle+
\omega^{5j}|2\rangle|5\rangle+\omega^{6j}|10\rangle|6\rangle\\
\hspace{1.1cm}
+\omega^{7j}|6\rangle|7\rangle+\omega^{8j}|5\rangle|8\rangle+\omega^{9j}|11\rangle|9\rangle+\omega^{10j}|0\rangle|10\rangle+\omega^{11j}|4\rangle|11\rangle) \\
A^5_{2,j}=\frac{1}{2\sqrt{3}}(|2\rangle |0\rangle +\omega^j|8\rangle|1\rangle+\omega^{2j}|10\rangle|2\rangle+\omega^{3j}|4\rangle|3\rangle+\omega^{4j}|9\rangle|4\rangle+\omega^{5j}|3\rangle|5\rangle
+\omega^{6j}|11\rangle|6\rangle\\
\hspace{1.1cm} +\omega^{7j}|7\rangle|7\rangle+\omega^{8j}|0\rangle|8\rangle
+\omega^{9j}|6\rangle|9\rangle+\omega^{10j}|1\rangle|10\rangle+\omega^{11j}|5\rangle|11\rangle) \\
A^5_{3,j}=\frac{1}{2\sqrt{3}}(|3\rangle |0\rangle +\omega^j|9\rangle|1\rangle+\omega^{2j}|11\rangle|2\rangle+\omega^{3j}|5\rangle|3\rangle+\omega^{4j}|10\rangle|4\rangle+\omega^{5j}|4\rangle|5\rangle
+\omega^{6j}|6\rangle|6\rangle\\
\hspace{1.1cm} +\omega^{7j}|8\rangle|1\rangle+\omega^{8j}|1\rangle|8\rangle
+\omega^{9j}|7\rangle|9\rangle+\omega^{10j}|2\rangle|10\rangle+\omega^{11j}|0\rangle|11\rangle) \\
A^5_{4,j}=\frac{1}{2\sqrt{3}}(|4\rangle |0\rangle +\omega^j|10\rangle|1\rangle+\omega^{2j}|6\rangle|2\rangle+\omega^{3j}|0\rangle|3\rangle+\omega^{4j}|11\rangle|4\rangle+\omega^{5j}|5\rangle|5\rangle
+\omega^{6j}|7\rangle|6\rangle\\
\hspace{1.1cm} +\omega^{7j}|9\rangle|7\rangle+\omega^{8j}|2\rangle|8\rangle
+\omega^{9j}|8\rangle|9\rangle+\omega^{10j}|3\rangle|10\rangle+\omega^{11j}|1\rangle|11\rangle) \\
A^5_{5,j}=\frac{1}{2\sqrt{3}}(|5\rangle |0\rangle +\omega^j|11\rangle|1\rangle+\omega^{2j}|7\rangle|2\rangle+\omega^{3j}|1\rangle|3\rangle+\omega^{4j}|6\rangle|4\rangle+\omega^{5j}|0\rangle|5\rangle
+\omega^{6j}|8\rangle|6\rangle\\
\hspace{1.1cm} +\omega^{7j}|10\rangle|7\rangle+\omega^{8j}|3\rangle|8\rangle
+\omega^{9j}|9\rangle|9\rangle+\omega^{10j}|4\rangle|10\rangle+\omega^{11j}|2\rangle|11\rangle) \\
A^5_{6,j}=\frac{1}{2\sqrt{3}}(|6\rangle |0\rangle +\omega^j|0\rangle|1\rangle+\omega^{2j}|2\rangle|2\rangle+\omega^{3j}|8\rangle|3\rangle+\omega^{4j}|1\rangle|4\rangle+\omega^{5j}|7\rangle|5\rangle
+\omega^{6j}|3\rangle|6\rangle\\
\hspace{1.1cm} +\omega^{7j}|5\rangle|7\rangle+\omega^{8j}|10\rangle|8\rangle
+\omega^{9j}|4\rangle|9\rangle+\omega^{10j}|11\rangle|10\rangle+\omega^{11j}|9\rangle|11\rangle) \\
A^5_{7,j}=\frac{1}{2\sqrt{3}}(|7\rangle |0\rangle +\omega^j|1\rangle|1\rangle+\omega^{2j}|3\rangle|2\rangle+\omega^{3j}|9\rangle|3\rangle+\omega^{4j}|2\rangle|4\rangle+\omega^{5j}|8\rangle|5\rangle
+\omega^{6j}|4\rangle|6\rangle\\
\hspace{1.1cm} +\omega^{7j}|0\rangle|7\rangle+\omega^{8j}|11\rangle|8\rangle
+\omega^{9j}|5\rangle|9\rangle+\omega^{10j}|6\rangle|10\rangle+\omega^{11j}|10\rangle|11\rangle) \\
A^5_{8,j}=\frac{1}{2\sqrt{3}}(|8\rangle |0\rangle +\omega^j|2\rangle|1\rangle+\omega^{2j}|4\rangle|2\rangle+\omega^{3j}|10\rangle|3\rangle+\omega^{4j}|3\rangle|4\rangle+\omega^{5j}|9\rangle|5\rangle
+\omega^{6j}|5\rangle|6\rangle\\
\hspace{1.1cm} +\omega^{7j}|1\rangle|7\rangle+\omega^{8j}|6\rangle|8\rangle
+\omega^{9j}|0\rangle|9\rangle+\omega^{10j}|7\rangle|10\rangle+\omega^{11j}|11\rangle|11\rangle) \\
A^5_{9,j}=\frac{1}{2\sqrt{3}}(|9\rangle |0\rangle +\omega^j|3\rangle|1\rangle+\omega^{2j}|5\rangle|2\rangle+\omega^{3j}|11\rangle|3\rangle+\omega^{4j}|4\rangle|4\rangle+\omega^{5j}|10\rangle|5\rangle
+\omega^{6j}|0\rangle|6\rangle\\
\hspace{1.1cm} +\omega^{7j}|2\rangle|7\rangle+\omega^{8j}|7\rangle|8\rangle
+\omega^{9j}|1\rangle|9\rangle+\omega^{10j}|8\rangle|10\rangle+\omega^{11j}|6\rangle|11\rangle) \\
A^5_{10,j}=\frac{1}{2\sqrt{3}}(|10\rangle |0\rangle +\omega^j|4\rangle|1\rangle+\omega^{2j}|0\rangle|2\rangle+\omega^{3j}|6\rangle|3\rangle+\omega^{4j}|5\rangle|4\rangle+\omega^{5j}|11\rangle|5\rangle
+\omega^{6j}|1\rangle|6\rangle\\
\hspace{1.1cm} +\omega^{7j}|3\rangle|7\rangle+\omega^{8j}|8\rangle|8\rangle
+\omega^{9j}|2\rangle|9\rangle+\omega^{10j}|9\rangle|10\rangle+\omega^{11j}|7\rangle|11\rangle) \\
A^5_{11,j}=\frac{1}{2\sqrt{3}}(|11\rangle |0\rangle +\omega^j|5\rangle|1\rangle+\omega^{2j}|1\rangle|2\rangle+\omega^{3j}|7\rangle|3\rangle+\omega^{4j}|0\rangle|4\rangle+\omega^{5j}|6\rangle|5\rangle
+\omega^{6j}|2\rangle|6\rangle\\
\hspace{1.1cm} +\omega^{7j}|4\rangle|7\rangle+\omega^{8j}|9\rangle|8\rangle
+\omega^{9j}|3\rangle|9\rangle+\omega^{10j}|10\rangle|10\rangle+\omega^{11j}|8\rangle|11\rangle) \\
     \end{array}
       \right.$
\end{center}
where $0\leq j\leq 11$. It is easy to check that the above six bases are  mutually unbiased.

\section{Example \ref{36}: MEBs and PB in $\mathbb{C}^{3}\otimes \mathbb{C}^{6}$}
\label{sec:AppC}

A $(3,5,2)$-DM without one row all being 0s on $\mathbb{Z}_3$ exists as follows.
\begin{equation}
\renewcommand\arraystretch{0.8}
M=\left(
\begin{array}{cccccc}
0&0&1&1&2&2\\
0&1&0&2&2&1\\
0&1&2&0&1&2\\
0&2&1&2&1&0\\
0&2&2&1&0&1\\
\end{array}
\right).
\end{equation}

Let $H^1_i=\hat{F}_3$, $H^2_j=\hat{F}_2$, $1\leq i\leq 4$, $0\leq j\leq 4$. Denote $S^r$ the development of the $r$th-row of $M$ for $0\leq r\leq 4$:
\begin{eqnarray*}
\qihao
\begin{small}
 \setlength{\arraycolsep}{2.2 pt}
 \renewcommand\arraystretch{0.8}
\begin{array}{lc}
\mbox{}&
\begin{array}{cccccc}0&1&2&3&4&5 \end{array}\\
\begin{array}{c}0\\1\\2 \end{array}&
\begin{array}{|c|c|c|c|c|c|}
\hline
0&0&1&1&2&2 \\
\hline
1&1&2&2&0&0 \\
\hline
2&2&0&0&1&1 \\
\hline
\end{array}
\end{array}~~
\begin{array}{lc}
\mbox{}&
\begin{array}{cccccc}0&1&2&3&4&5  \end{array}\\
\begin{array}{c}0\\1\\2 \end{array}&
\begin{array}{|c|c|c|c|c|c|}
\hline
0&1&0&2&2&1 \\
\hline
1&2&1&0&0&2\\
\hline
2&0&2&1&1&0 \\
\hline
\end{array}
\end{array}~~
\begin{array}{lc}
\mbox{}&
\begin{array}{cccccc}0&1&2&3&4&5  \end{array}\\
\begin{array}{c}0\\1\\2 \end{array}&
\begin{array}{|c|c|c|c|c|c|}
\hline
0&1&2&0&1&2 \\
\hline
1&2&0&1&2&0\\
\hline
2&0&1&2&0&1\\
\hline
\end{array}
\end{array}~~
\begin{array}{lc}
\mbox{}&
\begin{array}{cccccc}0&1&2&3&4&5  \end{array}\\
\begin{array}{c}0\\1\\2 \end{array}&
\begin{array}{|c|c|c|c|c|c|}
\hline
0&2&1&2&1&0 \\
\hline
1&0&2&0&2&1\\
\hline
2&1&0&1&0&2\\
\hline
\end{array}
\end{array}~~
\begin{array}{lc}
\mbox{}&
\begin{array}{cccccc}0&1&2&3&4&5  \end{array}\\
\begin{array}{c}0\\1\\2 \end{array}&
\begin{array}{|c|c|c|c|c|c|}
\hline
0&2&2&1&0&1 \\
\hline
1&0&0&2&1&2\\
\hline
2&1&1&0&2&0\\
\hline
\end{array}
\end{array}
\end{small}
\end{eqnarray*}
\hspace{2.1cm} $S^0$ \hspace{2.3cm}  $S^1$\hspace{2.5cm}  $S^2$\hspace{2.4cm} $S^3$ \hspace{2.4cm} $S^4$\newpage

Then the five bases can be written as,
\vspace{0.2cm}
\begin{center}
\small
 \setlength{\arraycolsep}{0.0 pt}
 \renewcommand\arraystretch{0.8}
  $  \left\{
          \begin{array}{ll}
A^0_{0,0,l}=\frac{1}{\sqrt{2}}|0\rangle (|0\rangle\pm|1\rangle)\\
A^0_{0,1,l}=\frac{1}{\sqrt{2}}|1\rangle (|2\rangle\pm|3\rangle)\\
A^0_{0,2,l}=\frac{1}{\sqrt{2}}|2\rangle (|4\rangle\pm|5\rangle)\\
A^0_{1,0,l}=\frac{1}{\sqrt{2}}|0\rangle (|4\rangle\pm|5\rangle)\\
A^0_{1,1,l}=\frac{1}{\sqrt{2}}|1\rangle (|0\rangle\pm|1\rangle)\\
A^0_{1,2,l}=\frac{1}{\sqrt{2}}|2\rangle (|2\rangle\pm|3\rangle)\\
A^0_{2,0,l}=\frac{1}{\sqrt{2}}|0\rangle (|2\rangle\pm|3\rangle)\\
A^0_{2,1,l}=\frac{1}{\sqrt{2}}|1\rangle (|4\rangle\pm|5\rangle)\\
A^0_{2,2,l}=\frac{1}{\sqrt{2}}|2\rangle (|0\rangle\pm|1\rangle)
     \end{array}
       \right.$
\end{center}
\begin{center}
\small
 \setlength{\arraycolsep}{0.0 pt}
 \renewcommand\arraystretch{0.8}
  $  \left\{
          \begin{array}{ll}
A^1_{0,j,l}=\frac{1}{\sqrt{6}}|0\rangle (|0\rangle\pm|2\rangle)+\omega^j|1\rangle (|1\rangle\pm|5\rangle)+\omega^{2j}|2\rangle (|3\rangle\pm|4\rangle)\\
A^1_{1,j,l}=\frac{1}{\sqrt{6}}|0\rangle (|3\rangle\pm|4\rangle)+\omega^j|1\rangle (|0\rangle\pm|2\rangle)+\omega^{2j}|2\rangle (|1\rangle\pm|5\rangle)\\
A^1_{2,j,l}=\frac{1}{\sqrt{6}}|0\rangle (|1\rangle\pm|5\rangle)+\omega^j|1\rangle (|3\rangle\pm|4\rangle)+\omega^{2j}|2\rangle (|0\rangle\pm|2\rangle)
     \end{array}
       \right.$
\end{center}
\begin{center}
\small
 \setlength{\arraycolsep}{0.0 pt}
 \renewcommand\arraystretch{0.8}
  $  \left\{
          \begin{array}{ll}
A^2_{0,j,l}=\frac{1}{\sqrt{6}}|0\rangle (|0\rangle\pm|3\rangle)+\omega^j|1\rangle (|1\rangle\pm|4\rangle)+\omega^{2j}|2\rangle (|2\rangle\pm|5\rangle)\\
A^2_{1,j,l}=\frac{1}{\sqrt{6}}|0\rangle (|2\rangle\pm|5\rangle)+\omega^j|1\rangle (|0\rangle\pm|3\rangle)+\omega^{2j}|2\rangle (|1\rangle\pm|4\rangle)\\
A^2_{2,j,l}=\frac{1}{\sqrt{6}}|0\rangle (|1\rangle\pm|4\rangle)+\omega^j|1\rangle (|2\rangle\pm|5\rangle)+\omega^{2j}|2\rangle (|0\rangle\pm|3\rangle)
     \end{array}
       \right.$
\end{center}
\begin{center}
\small
 \setlength{\arraycolsep}{0.0 pt}
 \renewcommand\arraystretch{0.8}
  $  \left\{
          \begin{array}{ll}
A^3_{0,j,l}=\frac{1}{\sqrt{6}}|0\rangle (|0\rangle\pm|5\rangle)+\omega^j|1\rangle (|2\rangle\pm|4\rangle)+\omega^{2j}|2\rangle (|1\rangle\pm|3\rangle)\\
A^3_{1,j,l}=\frac{1}{\sqrt{6}}|0\rangle (|1\rangle\pm|3\rangle)+\omega^j|1\rangle (|0\rangle\pm|5\rangle)+\omega^{2j}|2\rangle (|2\rangle\pm|4\rangle)\\
A^3_{2,j,l}=\frac{1}{\sqrt{6}}|0\rangle (|2\rangle\pm|4\rangle)+\omega^j|1\rangle (|1\rangle\pm|3\rangle)+\omega^{2j}|2\rangle (|0\rangle\pm|5\rangle)
     \end{array}
       \right.$
\end{center}
\begin{center}
\small
 \setlength{\arraycolsep}{0.0 pt}
 \renewcommand\arraystretch{0.8}
  $  \left\{
          \begin{array}{ll}
A^4_{0,j,l}=\frac{1}{\sqrt{6}}|0\rangle (|0\rangle\pm|4\rangle)+\omega^j|1\rangle (|3\rangle\pm|5\rangle)+\omega^{2j}|2\rangle (|1\rangle\pm|2\rangle)\\
A^4_{1,j,l}=\frac{1}{\sqrt{6}}|0\rangle (|1\rangle\pm|2\rangle)+\omega^j|1\rangle (|0\rangle\pm|4\rangle)+\omega^{2j}|2\rangle (|3\rangle\pm|5\rangle)\\
A^4_{2,j,l}=\frac{1}{\sqrt{6}}|0\rangle (|3\rangle\pm|5\rangle)+\omega^j|1\rangle (|1\rangle\pm|2\rangle)+\omega^{2j}|2\rangle (|0\rangle\pm|4\rangle)
     \end{array}
       \right.$
\end{center}
where $0\leq j\leq 2$, $0\leq l\leq 1$ and $\omega=e^{\frac{2\pi \sqrt{-1}}{3}}$.

\section{Example \ref{39}: MUMEBs and PB in $\mathbb{C}^{3}\otimes \mathbb{C}^{9}$}
\label{sec:AppD}

A $(3,4,3)$-DM exists on $\mathbb{Z}_3$ as follows:
\begin{equation}
M=\left(
\begin{array}{cccccccccccc}
0&0&0&1&1&1&2&2&2\\
0&1&2&0&1&2&0&1&2\\
0&1&2&1&2&0&2&0&1\\
0&2&1&1&0&2&2&1&0\\
\end{array}
\right).\label{3}
\end{equation}
Let $H^1_i$, $H^2_j$ be the following complex Hadamard matrices for $1\leq i\leq 3$, $0\leq j\leq 3$,
\vspace{-0.4cm}\begin{small}\begin{center}{$H^1_1=D\hat{F}_3= \left( {\begin{array}{*{20}{c}}
1&1&1\\
\omega&\omega^2&1\\
\omega&1&\omega^2\\
\end{array}} \right)$,~~ $H^1_2=D^2\hat{F}_3 =\left( {\begin{array}{*{20}{c}}
1&1&1\\
\omega^2&1&\omega\\
\omega^2&\omega&1\\
\end{array}} \right)$},
\end{center}
\end{small}
\hspace{4cm}$H^2_j=H^1_3=\hat{F}_3= \left( {\begin{array}{*{20}{c}}
1&1&1\\
1&\omega&\omega^2\\
1&\omega^2&\omega\\
\end{array}} \right),$

\noindent where $D=diag(1,\omega,\omega), \omega=e^{\frac{2\pi \sqrt{-1}}{3}}$.

Denote the development of the $r$th-row as $S^{r}$
 for $0\leq r\leq 3$,
\begin{eqnarray*}
\qihao
\begin{small}
 \setlength{\arraycolsep}{2 pt}
\begin{array}{lc}
\mbox{}&
\begin{array}{ccccccccc}0&1&2&3&4&5&6&7&8  \end{array}\\
\begin{array}{c}0\\1\\2 \end{array}&
\begin{array}{|c|c|c|c|c|c|c|c|c|}
\hline
0&0&0&1&1&1&2&2&2 \\
\hline
1&1&1&2&2&2&0&0&0 \\
\hline
2&2&2&0&0&0&1&1&1 \\
\hline
\end{array}
\end{array}~~
\begin{array}{lc}
\mbox{}&
\begin{array}{ccccccccc}0&1&2&3&4&5&6&7&8  \end{array}\\
\begin{array}{c}0\\1\\2 \end{array}&
\begin{array}{|c|c|c|c|c|c|c|c|c|}
\hline
0&1&2&0&1&2&0&1&2 \\
\hline
1&2&0&1&2&0&1&2&0 \\
\hline
2&0&1&2&0&1&2&0&1 \\
\hline
\end{array}
\end{array}~~
\begin{array}{lc}
\mbox{}&
\begin{array}{ccccccccc}0&1&2&3&4&5&6&7&8  \end{array}\\
\begin{array}{c}0\\1\\2 \end{array}&
\begin{array}{|c|c|c|c|c|c|c|c|c|}
\hline
0&1&2&1&2&0&2&0&1 \\
\hline
1&2&0&2&0&1&0&1&2\\
\hline
2&0&1&0&1&2&1&2&0\\
\hline
\end{array}
\end{array}~~
\begin{array}{lc}
\mbox{}&
\begin{array}{ccccccccc}0&1&2&3&4&5&6&7&8  \end{array}\\
\begin{array}{c}0\\1\\2 \end{array}&
\begin{array}{|c|c|c|c|c|c|c|c|c|}
\hline
0&2&1&1&0&2&2&1&0 \\
\hline
1&0&2&2&1&0&0&2&1\\
\hline
2&1&0&0&2&1&1&0&2\\
\hline
\end{array}
\end{array}
\end{small}
\end{eqnarray*}
\hspace{2.2cm} $S^0$ \hspace{3.2cm}  $S^1$\hspace{3.5cm}  $S^2$\hspace{3.3cm} $S^3$\\
\quad \quad The four bases can be written as
\begin{center}
\small
 \setlength{\arraycolsep}{0.0 pt}
  $  \left\{
          \begin{array}{ll}
A^0_{0,0,l}=\frac{1}{\sqrt{3}}|0\rangle (|0\rangle+\omega^l|1\rangle+\omega^{2l}|2\rangle)\\
A^0_{0,1,l}=\frac{1}{\sqrt{3}}|1\rangle (|3\rangle+\omega^l|4\rangle+\omega^{2l}|5\rangle)\\
A^0_{0,2,l}=\frac{1}{\sqrt{3}}|2\rangle (|6\rangle+\omega^l|7\rangle+\omega^{2l}|8\rangle)\\
A^0_{1,0,l}=\frac{1}{\sqrt{3}}|0\rangle (|6\rangle+\omega^l|7\rangle+\omega^{2l}|8\rangle)\\
A^0_{1,1,l}=\frac{1}{\sqrt{3}}|1\rangle (|0\rangle+\omega^l|1\rangle+\omega^{2l}|2\rangle)\\
A^0_{1,2,l}=\frac{1}{\sqrt{3}}|2\rangle (|3\rangle+\omega^l|4\rangle+\omega^{2l}|5\rangle)\\
A^0_{2,0,l}=\frac{1}{\sqrt{3}}|0\rangle (|3\rangle+\omega^l|4\rangle+\omega^{2l}|5\rangle)\\
A^0_{2,1,l}=\frac{1}{\sqrt{3}}|1\rangle (|6\rangle+\omega^l|7\rangle+\omega^{2l}|8\rangle)\\
A^0_{2,2,l}=\frac{1}{\sqrt{3}}|2\rangle (|0\rangle+\omega^l|1\rangle+\omega^{2l}|2\rangle)
     \end{array}
       \right.$
\end{center}
\begin{center}
\small
 \setlength{\arraycolsep}{0.0 pt}
  $  \left\{
          \begin{array}{ll}
A^1_{0,j,l}=\frac{1}{3}[|0\rangle(|0\rangle+\omega^l|3\rangle+\omega^{2l}|6\rangle)
+\omega^{j+1}|1\rangle (|1\rangle+\omega^l|4\rangle+\omega^{2l}|7\rangle)\\
\hspace{1.3cm}+\omega^{2j+1}|2\rangle (|2\rangle+\omega^l|5\rangle+\omega^{2l}|8\rangle)] \\
A^1_{1,j,l}=\frac{1}{3}[|0\rangle(|2\rangle+\omega^l|5\rangle+\omega^{2l}|8\rangle)
+\omega^{j+1}|1\rangle (|0\rangle+\omega^l|3\rangle+\omega^{2l}|6\rangle)\\
\hspace{1.3cm}
+\omega^{2j+1}|2\rangle (|1\rangle+\omega^l|4\rangle+\omega^{2l}|7\rangle)] \\
A^1_{2,j,l}=\frac{1}{3}[|0\rangle(|1\rangle+\omega^l|4\rangle+\omega^{2l}|7\rangle)
+\omega^{j+1}|1\rangle (|2\rangle+\omega^l|5\rangle+\omega^{2l}|8\rangle)\\
\hspace{1.3cm}
+\omega^{2j+1}|2\rangle (|0\rangle+\omega^l|3\rangle+\omega^{2l}|6\rangle)] \\
     \end{array}
       \right.$
\end{center}\newpage
\begin{center}
\small
 \setlength{\arraycolsep}{0.0 pt}
  $  \left\{
          \begin{array}{ll}
A^2_{0,j,l}=\frac{1}{3}[|0\rangle(|0\rangle+\omega^l|5\rangle+\omega^{2l}|7\rangle)
+\omega^{j+2}|1\rangle (|1\rangle+\omega^l|3\rangle+\omega^{2l}|8\rangle)\\
\hspace{1.3cm}+\omega^{2j+2}|2\rangle (|2\rangle+\omega^l|4\rangle+\omega^{2l}|6\rangle)] \\
A^2_{1,j,l}=\frac{1}{3}[|0\rangle(|2\rangle+\omega^l|4\rangle+\omega^{2l}|6\rangle)
+\omega^{j+2}|1\rangle (|0\rangle+\omega^l|5\rangle+\omega^{2l}|7\rangle)\\
\hspace{1.3cm}
+\omega^{2j+2}|2\rangle (|1\rangle+\omega^l|3\rangle+\omega^{2l}|8\rangle)\\
A^2_{2,j,l}=\frac{1}{3}[|0\rangle(|1\rangle+\omega^l|3\rangle+\omega^{2l}|8\rangle)
+\omega^{j+2}|1\rangle (|2\rangle+\omega^l|4\rangle+\omega^{2l}|6\rangle)\\
\hspace{1.3cm}
+\omega^{2j+2}|2\rangle (|0\rangle+\omega^l|5\rangle+\omega^{2l}|7\rangle)] \\
     \end{array}
       \right.$
\end{center}
\begin{center}
\small
  \hspace{-0.4cm}\setlength{\arraycolsep}{0.0 pt}
  $ \left\{
          \begin{array}{ll}
A^3_{0,j,l}=\frac{1}{3}[|0\rangle(|0\rangle+\omega^l|4\rangle+\omega^{2l}|8\rangle)
+\omega^{j}|1\rangle (|2\rangle+\omega^l|3\rangle+\omega^{2l}|7\rangle)\\
\hspace{1.3cm}+\omega^{2j}|2\rangle (|1\rangle+\omega^l|5\rangle+\omega^{2l}|6\rangle)] \\
A^3_{1,j,l}=\frac{1}{3}[|0\rangle(|1\rangle+\omega^l|5\rangle+\omega^{2l}|6\rangle)
+\omega^{j}|1\rangle (|0\rangle+\omega^l|4\rangle+\omega^{2l}|8\rangle)\\
\hspace{1.3cm}
+\omega^{2j}|2\rangle (|2\rangle+\omega^l|3\rangle+\omega^{2l}|7\rangle)] \\
A^3_{2,j,l}=\frac{1}{3}[|0\rangle(|2\rangle+\omega^l|3\rangle+\omega^{2l}|7\rangle)
+\omega^{j}|1\rangle (|1\rangle+\omega^l|5\rangle+\omega^{2l}|6\rangle)\\
\hspace{1.3cm}
+\omega^{2j}|2\rangle (|0\rangle+\omega^l|4\rangle+\omega^{2l}|8\rangle)] \\
     \end{array}
       \right.$
\end{center}
\noindent where $0\leq j,l\leq 2$. It is easy to check that the above four bases are mutually unbiased.

\end{document}